# Buffering Time Strategies for Wireless Full-duplex Systems under Poisson Traffic

Makoto Kobayashi, Shunsuke Saruwatari, and Takashi Watanabe

**Abstract**

Full-duplex wireless communication has the potential to double the capacity of wireless networks by reducing the band occupancy of transmissions. However, a full-duplex capability cannot always reduce the band occupancy because the real traffic is not fully buffered. Buffering time while waiting for a packet to arrive at an opposite node is expected to reduce the band occupancy. In this study, we provide the first theoretical analysis of band occupancy and the mean waiting time for full-duplex communication with and without buffering time under traffic that is not fully buffered based on queueing theory, as well as the closed-form results. We also present the results of simulations of band occupancy and the mean waiting time. The basic analysis provided in this study shows how the mean waiting time and band occupancy are affected by the buffering time. When the buffering time is half the packet length, the band occupancy is reduced by approximately 15 %. In addition, under asymmetrical traffic, the results suggest that the buffering time should not be set at the node who has a higher traffic intensity compared with another node. These results support the design of a full-duplex medium access control protocol and devices.

## I. Introduction

Full-duplex wireless communication is a key technology for enhancing the capacity of next-generation wireless networks. Recent developments in physical layer techniques for self-interference cancellation have transformed full-duplex wireless communication into a practical technology. In [1]–[3], the implementation of full-duplex wireless local area networks (LANs) was described.

The authors are with the Department of Information Networking, Graduate School of Information Science and Technology, Osaka University, Suita, Osaka, 565-0871 JAPAN (e-mail: {kobayshi.makoto, saru, watanabe}@ist.osaka-u.ac.jp).

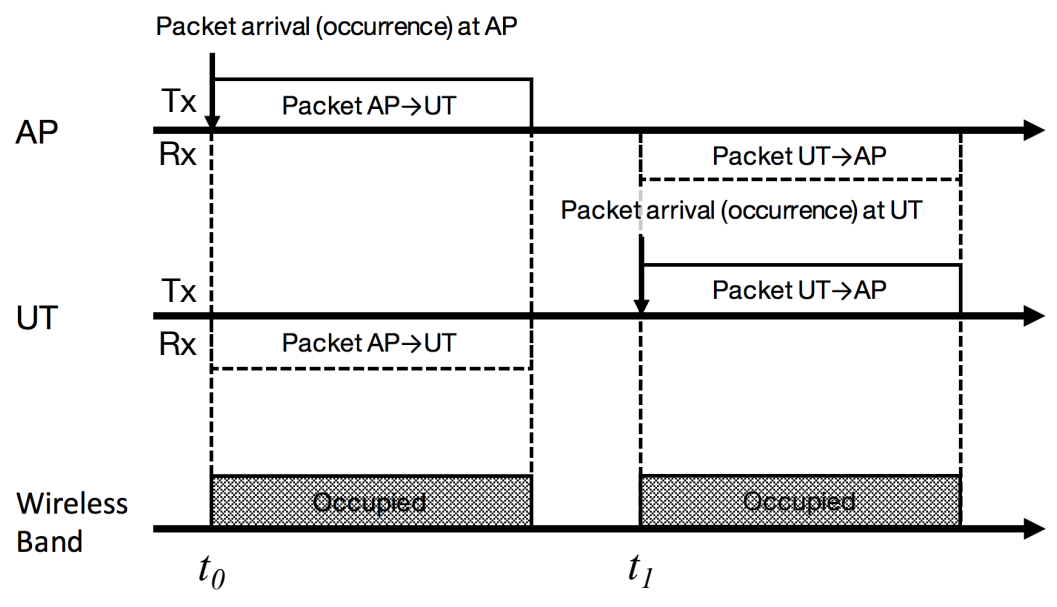

Fig. 1. Full-duplex wireless communication does not reduce the band occupancy.

According to [3], single antenna full-duplex communication achieves 110 dB self-interference cancellation and almost doubles the throughput compared with half-duplex communication. These implementations assume that two nodes simultaneously exchange frames in a channel with the same frequency. Theoretical analyses of the physical layer have explained the performance of other variations of full-duplex communication [4]–[6] such as three node full-duplexing.

In addition to full-duplex studies in the physical layer, extensive studies have considered the medium access control (MAC) protocol for full-duplex wireless LAN [7]–[19]. In [7]–[9], several full-duplex MAC protocols were described to improve the throughput of ad-hoc wireless communication. Full-duplex MAC protocols for infrastructure networks have also been studied [10]–[19]. In [10]–[17], a full-duplex MAC protocol was proposed for throughput enhancement. In addition, an energy-efficient MAC protocol for a full-duplex network infrastructure was proposed [18], [19].

Previous studies have shown that full-duplex communications can enhance the capacity of the physical layer and MAC layer. However, in addition to the capacity, we must consider the band occupancy in full-duplex communications, which is important because recent wireless networks tend to share the same frequency channel.

Up-link and down-link traffic share a channel at the same time, so full-duplex communications have the potential to reduce the band occupancy. However, full-duplex capability cannot always reduce the band occupancy. Fig. 1 shows an example of a case where the full-duplex capability does not reduce the band occupancy. In Fig. 1, a down-link packet arrives (occurs) at a full-duplex capable access point (AP) at $t_0$ and the access point sends the packet to a full-duplex capable user terminal (UT) immediately. The uplink packet then arrives (occurs) at the user terminal at $t_1$ and the user terminal also sends the packet to the access point immediately. In this case, access point and user terminal transmit data and occupy a wireless band, respectively. Therefore, the band occupancy in Fig. 1 is equal to that of half-duplex communication even if the access point and user terminal have full-duplex capability. Now, we consider the case where the access point holds transmission over until $t_1$ (as depicted in Fig. 2). We refer to the duration of holding packets as the buffering time. After the buffering time, access point and user terminal transmit a packet at the same time. The buffering time compresses the band occupancy to one-second of the band occupancy in half-duplex communication. However, the buffering time increases the waiting time for each packet, which causes a delay.

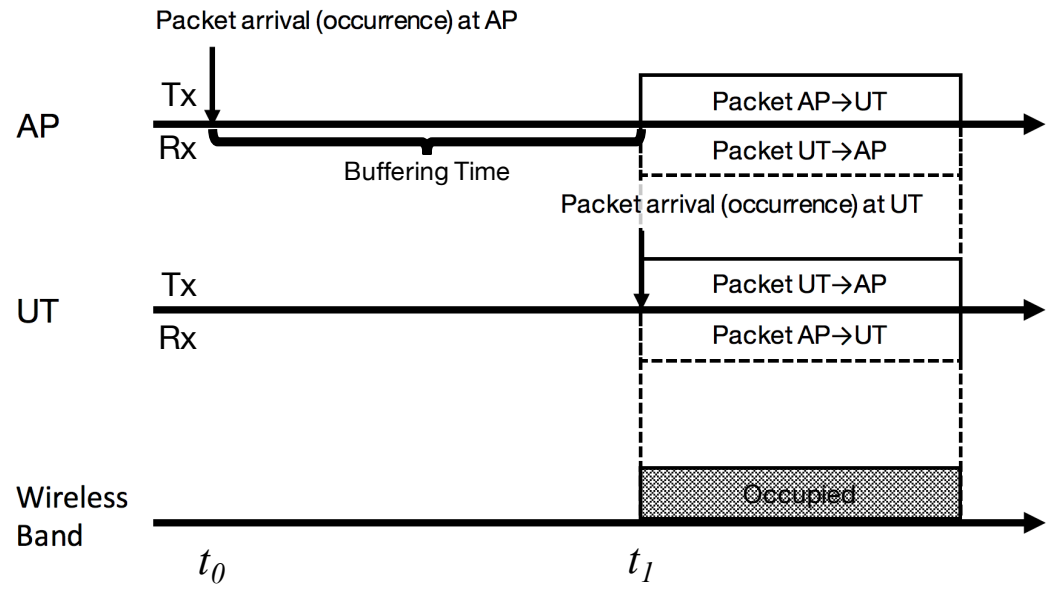

Fig. 2. Buffering time enables full-duplex to reduce the band occupancy.

In this study, we present a theoretical analysis of band occupancy and the waiting time for full-duplex communication with buffering time. We analyze full-duplex communication under Poisson traffic and deterministic service duration (packet length) using queueing theory and simulations. In addition, we analyze the effect of the buffering time on full-duplex communication. To exploit the full-duplex capability in practical networks, the access point and any user terminal need to have packets to send at the same time. Thus, a buffering time while waiting for a packet's arrival (occurrence) at the opposite node(s) is expected to enhance full-duplex performance, i.e., when an access point has a packet(s) to send, the access point waits for a packet's arrival at any user terminal.

The main contributions of this study are summarized as follows.

- We analyze the potential for full-duplex communication under Poisson traffic that is not fully buffered with a deterministic service duration (packet length) for each access point and user terminal.
- This is the first theoretical analysis using queueing theory of full-duplex communication under traffic that is not fully buffered. We theoretically determine the upper and lower bounds of band occupancy and the mean waiting time. The band occupancy is the proportion of the transmission's duration relative to the total time. The waiting time for a packet is the duration between the arrival of the packet and the end of the packet's transmission.

- The effect of the buffering time on full-duplex communication under Poisson traffic is shown in this study. The closed-form results presented in this study demonstrate the criteria needed to decide the length of the buffering time. The results suggest that the optimal buffering time depends on the packet arrival rate, the memory size of each node, and traffic asymmetry.

The remainder of this paper is organized as follows. In Section II, we present the system models for half-duplex communication, ideal full-duplex communication, practical full-duplex communication without buffering time, and practical full-duplex communication with buffering time. In Section III, we consider the band occupancy of full-duplex communication based on queueing theory. In Section IV, we determine the mean packet waiting time in full-duplex communication according to queueing theory. In Section V, we present the results of our performance analysis based on queuing theory and simulations. Finally, we give our conclusions in Section VI.

## II. System Model

Consider a pair of full-duplex transceivers: an access point and user terminal. We assume that the traffic model is a Poisson packet arrival process. We denote the traffic arrival rates at access point (down-link) and user terminal (up-link) by $\lambda_{\mathrm{AP}}$ and $\lambda_{\mathrm{UT}}$, respectively. Access point and user terminal serve (send) their own packets on a "first come, first served" basis. We set the packet lengths (service times) for access point and user terminal as deterministic time $b_{\mathrm{AP}}$ and $b_{\mathrm{UT}}$, respectively. We define $b_{\mathrm{AP}} = b_{\mathrm{UT}} = b$, because the packet size is usually static length in the MAC layer. For normalization, we set the packet length as a dimensionless quantity of 1 ($b = 1$) and the traffic arrival rates are normalized based on the packet length. The traffic intensities are defined as $\rho_{\mathrm{AP}} = \lambda_{\mathrm{AP}} b$, $\rho_{\mathrm{UT}} = \lambda_{\mathrm{UT}} b$. We assume that $\rho_{\mathrm{AP}} < 1$ and $\rho_{\mathrm{UT}} < 1$. Our analysis considers a wireless network in a steady state.

Note that this assumption can be extended to the multi-user terminal case if we assume that $\lambda_{\mathrm{UT}}$ is the sum of the traffic arrival rate at user terminals. In the present study, we compare the performance of the following four systems to evaluate the effects of the buffering time on full-duplex communication.

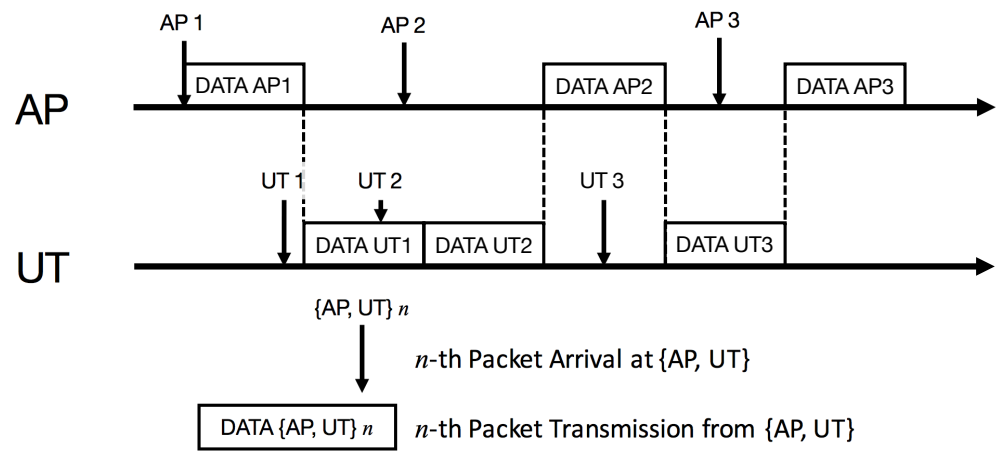

Fig. 3. Half-duplex Communication

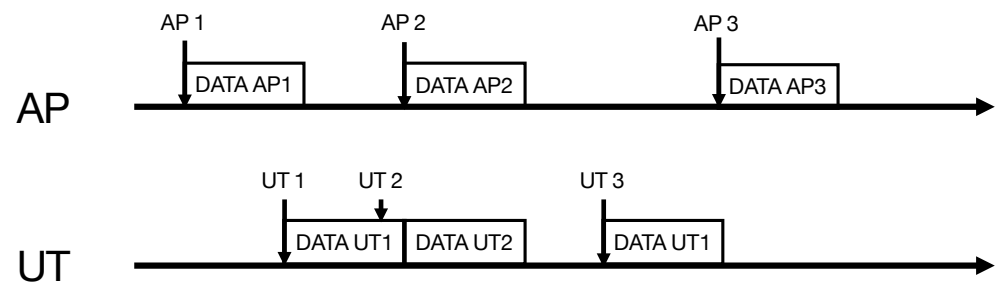

Fig. 4. Ideal Full-duplex Communication

1) Half-duplex Communication

   Half-duplex communication is a current wireless communication system. We simplify half-duplex communication and Fig. 3 shows the type of half-duplex communication considered in this study. In half-duplex communication, the access point and user terminals transmit packets time divisionally. The first packets to arrive at the access point or user terminals will be served first. For simplicity, DATA in Fig. 3 comprises a protocol sequence for data frame transmission in the IEEE 802.11 MAC protocol.

2) Ideal Full-duplex Communication

   Ideal full-duplex communication is full-duplex communication without any restrictions on simultaneous transmission. Fig. 4 shows an example of ideal full-duplex communication. In ideal full-duplex communication, each access point and user terminal starts data transmission at any time they want to send a packet. However, ideal full-duplex communication is not practical. The reason is that, in practical full-duplex communication, access point and user terminal exchange preamble packets to allow physical layer self-interference cancelation [1]–[3]. Therefore, access point and user terminal need to start full-duplex communication at the same time. We refer to this as simultaneous transmission restriction.

3) Practical Full-duplex Communication without Buffering Time

   Practical full-duplex communication satisfies the simultaneous transmission restriction. The simultaneous transmission restriction requires that access point and user terminal start transmission at the same time in full-duplex communication. Algorithm 1 shows the operation of the access point (user terminal) in practical full-duplex communication without buffering time. Note that access point and user terminals operate according to the same algorithm. Fig. 5 shows an example of full-duplex communication. In Fig. 5, "DATA UT 1" arrives at the user terminal during access point transmission. The user terminal then waits for the end of access point transmission due to the simultaneous transmission restriction.

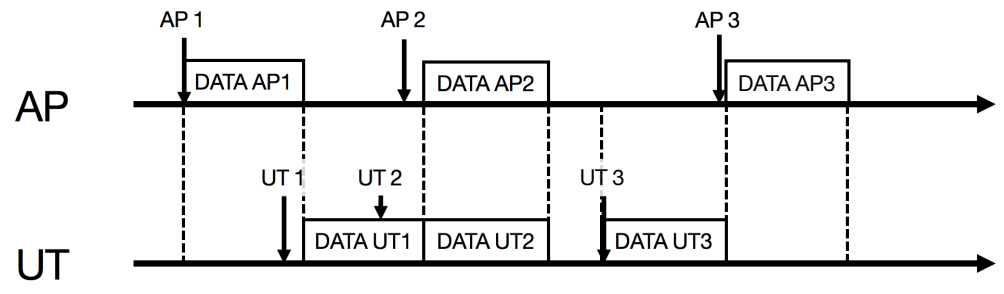

Fig. 5. Practical Full-duplex without Buffering Time

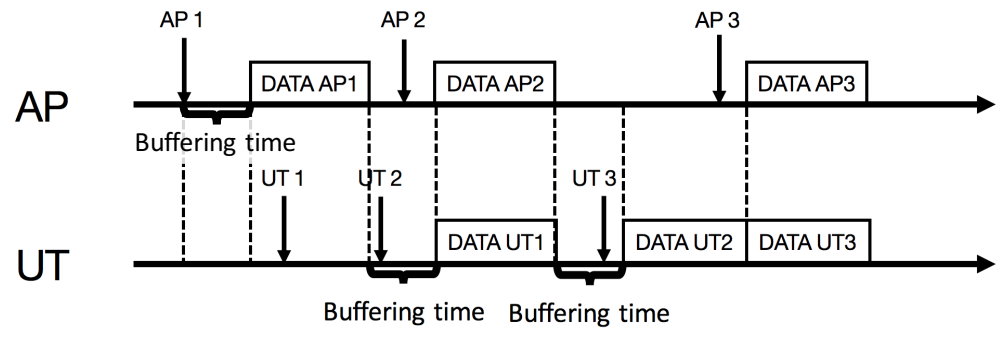

Fig. 6. Practical Full-duplex with Buffering Time

---

**Algorithm 1** Access point (user terminal) operation in full-duplex communication without buffering time

---

**while true do**
    **if** Have packet(s) **then**
        Wait while carrier busy
        Transmit packet to user terminal (access point)
    **end if**
**end while**

---

4) Practical Full-duplex Communication with Buffering Time

Now, we consider full-duplex communication with buffering time. Full-duplex communication with buffering time is expected to reduce the band occupancy. In the contrast to the band occupancy reduction, the buffering time increases the waiting time, i.e., the time duration between the arrival of a packet and the end of packet transmission. Thus, it is necessary to clarify the effects of buffering time. Fig. 6 shows an example of full-duplex communication with buffering time. In Fig. 6, user terminal sets the buffering time before sending "DATA UT1." Algorithm 2 shows the operation of the access point (user terminal) in full-duplex communication with buffering time. We assume that access point and user terminal set the buffering times as $\tau_{\mathrm{AP}}$ and $\tau_{\mathrm{UT}}$, respectively.

**Algorithm 2** Access point (user terminal) operation in full-duplex communication with buffering time

```
while true do
  if Have packet(s) then
    Wait for τ_AP (τ_UT)
    Wait while carrier busy
    Transmit packet to user terminal (access point)
  end if
end while
```

## III. Band Occupancy

Full-duplex capability is expected to reduce the band occupancy. In addition, the buffering time compresses the band occupancy further. In this section, we provide a theoretical analysis of band occupancy. First, the band occupancy of half-duplex communication is shown by Lemma 1. Second, the band occupancy of the ideal full-duplex communication is shown by Lemma 2. Third, the upper and lower bounds of band occupancy in practical full-duplex communication without buffering time are shown by Lemma 3. Finally, Theorems 1 and 2 show the upper and lower bounds of band occupancy in full-duplex communication with buffering time, respectively. Theorems 1 and 2 include Lemma 3 when the buffering time length is zero.

**Lemma 1 (Band Occupancy in Half-Duplex Communication):** *The band occupancy in half-duplex communication is*

$$\beta_{\mathrm{HD}} = \min\left(\rho_{\mathrm{AP}} + \rho_{\mathrm{UT}}, 1\right). \tag{1}$$

*Proof:* access point and user terminal occupy the band independently. The band occupancies in half-duplex communication for access point and user terminal are the same as each traffic intensity, $\rho_{\mathrm{AP}}$ and $\rho_{\mathrm{UT}}$, respectively. Therefore, the band occupancy in half-duplex communication is the sum of the traffic intensities $\rho_{\mathrm{AP}}$ and $\rho_{\mathrm{UT}}$ when $\rho_{\mathrm{AP}} + \rho_{\mathrm{UT}} \leq 1$ When the summed traffic intensity is greater than 1 ($\rho_{\mathrm{AP}} + \rho_{\mathrm{UT}} > 1$), the band occupancy in half-duplex communication is 1, even if half-duplex communication is not in a steady state. ■

**Lemma 2 (Band Occupancy in Ideal Full-Duplex Communication):** *The band occupancy in ideal full-duplex communication is*

$$\beta_{\mathrm{IFD}} = \rho_{\mathrm{AP}} + \rho_{\mathrm{UT}} - \rho_{\mathrm{AP}} \cdot \rho_{\mathrm{UT}}. \tag{2}$$

*Proof:* access point and user terminal transmit data independently in ideal full-duplex communication. The probability that access point does not use the band is $1-\rho_{\mathrm{AP}}$. The probability that user terminal does not use the band is also $1-\rho_{\mathrm{UT}}$ like that for access point. Hence, the band occupancy in ideal full-duplex communication is formulated as $\beta_{\mathrm{IFD}} = 1 - (1 - \rho_{\mathrm{AP}})(1 - \rho_{\mathrm{UT}})$. ■

Now, we consider the band occupancy in practical full-duplex communication without buffering time.

**Lemma 3 (Band Occupancy in Practical Full-Duplex Communication without Buffering Time):** *The band occupancy in practical full-duplex communication without buffering time ($\beta$) is bounded as follows.*

$$\beta < \beta_{\mathrm{HD}}(:= \beta_{\max}) \tag{3}$$

$$\beta > \beta_{\mathrm{IFD}}(:= \beta_{\min}) \tag{4}$$

*Proof:* We consider the arrival of the $n$-th packet at access point. Now, $p(n)$ is the probability that access point transmits the $n$-th packet by half-duplex communication. If both packets arrive at access point and user terminal during the transmission of the $(n-1)$-th access point packet, the $n$-th packet must be transmitted by full-duplex communication in practical full-duplex communication without buffering time where $n > 1$. Thus, $p(n)$ is bounded by $p(n) \leq 1 - \{1 - \exp(-\lambda_{\mathrm{AP}} b)\}\{1 - \exp(-\lambda_{\mathrm{UT}} b)\}$ when $n > 1$. The probability that all of the access point packets are transmitted by half-duplex communication is $\mathrm{P}\{\text{all packet HD}\} = \lim_{n\to\infty} \prod_{k=1}^{n} p(k) = 0$, where $p(1) = 1$. This probability indicates that at least one instance of full-duplex communication occurs in practical full-duplex communication without buffering time. Therefore, the band occupancy in practical full-duplex communication without buffering time is less than that in half-duplex communication, as shown in (3).

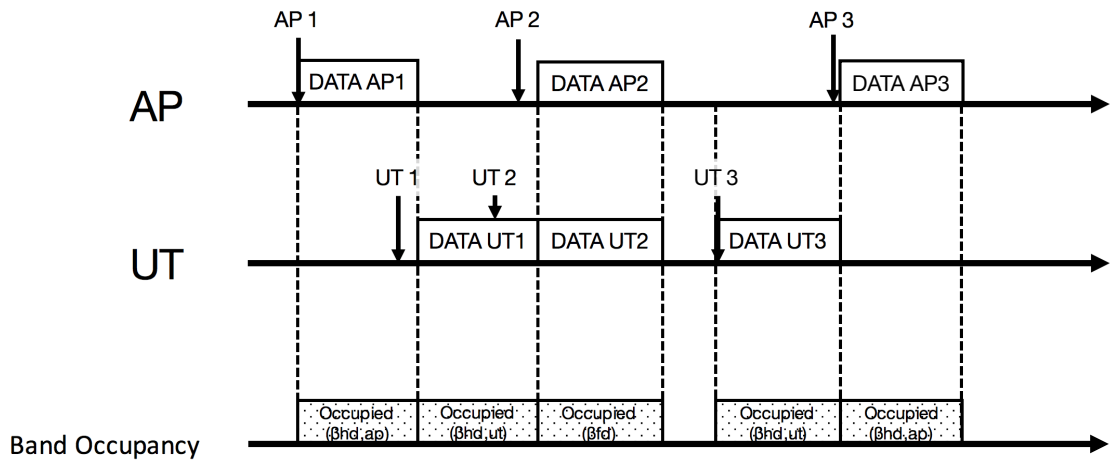

Fig. 7. Band Occupancy by Each Transmission

Next, we consider the simultaneous transmission restriction. We assume that the $n$-th access point packet arrives when the band is unoccupied. Note that the probability that the band is unoccupied when the $n$-th access point packet arrives is $(1-\beta)$, and it is larger than zero because of the "Poisson arrivals see time averages" (PASTA) rule [20]. In practical full-duplex communication without buffering time, the $n$-th access point packet is sent by half-duplex communication even if one or more packets arrive at user terminal during the transmission of the $n$-th access point packet. When packet(s) arrive at user terminal but not at access point during the transmission of the $n$-th access point packet, the band occupancy is increased compared with ideal full-duplex communication. The probability that packet(s) arrive at user terminal but not at access point during the transmission of the $n$-th access point packet is $\exp(-\lambda_{\mathrm{AP}}b)\left\{1-\exp(-\lambda_{\mathrm{UT}}b)\right\} > 0$. Therefore, the band occupancy in practical full-duplex communication is larger than that in ideal full-duplex communication, as shown in (4). ■

The previous Lemmas show the band occupancy of full-duplex communication without buffering time. Next, we consider the band occupancy of practical full-duplex communication with buffering time. The buffering time is expected to reduce the band occupancy. We clarify the effect of the buffering time as follows. Before showing the band occupancy under practical full-duplex communication with buffering time, we show the band occupancy by each transmission in practical full-duplex communication without buffering time. Fig. 7 shows the band occupancy of each transmission in practical full-duplex communication. For example, as shown in Fig. 7, the first half-duplex communication from access point occupies the band.

**Lemma 4 (Band Occupancy by Each Transmission in Practical Full-duplex Communication without Buffering Time):** *The band occupancies by each transmission in half-duplex communication from access point ($\beta_{\mathrm{AP}}(\beta)$), half-duplex communication from user terminal ($\beta_{\mathrm{UT}}(\beta)$),* *and full-duplex communication ($\beta_{\mathrm{FD}}(\beta)$) are*

$$\beta_{\mathrm{UT}}(\beta) = \beta - \rho_{\mathrm{AP}} \tag{5}$$

$$\beta_{\mathrm{FD}}(\beta) = \rho_{\mathrm{AP}} + \rho_{\mathrm{UT}} - \beta \tag{6}$$

$$\beta_{\mathrm{AP}}(\beta) = \beta - \rho_{\mathrm{UT}}, \tag{7}$$

*respectively, when we obtain the accurate band occupancy ($\beta$). We recall that the service length of full-duplex communication $b_{\mathrm{FD}} = \max(b_{\mathrm{AP}}, b_{\mathrm{UT}}) = b$.*

Thus, we can obtain the upper bound on band occupancy in practical full-duplex communication with buffering time.

**Theorem 1 (An Upper Bound on Band Occupancy in Practical Full-duplex Communication with Buffering Time):** *The band occupancy $\tilde{\beta}$ in practical full-duplex communication with buffering time is bounded by the following equations:*

$$\tilde{\beta} \leq \rho_{\mathrm{AP}} + \rho_{\mathrm{UT}} - \max\left[\rho_{\mathrm{AP}}\left\{1 - \exp\left(-\lambda_{\mathrm{UT}}\tau_{\mathrm{AP}}\right)\right\}, \rho_{\mathrm{UT}}\left\{1 - \exp\left(-\lambda_{\mathrm{AP}}\tau_{\mathrm{UT}}\right)\right\}\right] \qquad (\textit{for}\ \rho_{\mathrm{AP}} + \rho_{\mathrm{UT}} < 1)$$

$$\tilde{\beta} \leq \min\left[1 - (1 - \rho_{\mathrm{UT}})\left(1 - e^{-\lambda_{\mathrm{UT}}\tau_{\mathrm{AP}}}\right), 1 - (1 - \rho_{\mathrm{AP}})\left(1 - e^{-\lambda_{\mathrm{AP}}\tau_{\mathrm{UT}}}\right)\right] \qquad (\textit{for}\ \rho_{\mathrm{AP}} + \rho_{\mathrm{UT}} \geq 1).$$

*Proof:* We consider that one node does not set the buffering time.First, we assume that user terminal does not set the buffering time ($\tau_{\mathrm{UT}} = 0$). When the buffering time of access point is also zero ($\tau_{\mathrm{AP}} = 0$), $\frac{\beta_{\mathrm{AP}}(\beta)}{b}$ packets are transmitted by half-duplex transmission from access point. Access point attaches the buffering time to these packets to reduce half-duplex communication and the band occupancy. The probability that the packets with buffering time are finally transmitted by full-duplex transmission is $1 - \exp(-\lambda_{\mathrm{UT}}\tau_{\mathrm{AP}})$. This probability is equal to the probability that one or more packets arrives at user terminal in the access point buffering time ($\tau_{\mathrm{AP}}$). Thus, when the buffering time is $\tau_{\mathrm{AP}}$ ($> 0$), $\frac{\beta_{\mathrm{AP}}(\beta)}{b}\{1 - \exp(-\lambda_{\mathrm{UT}}\tau_{\mathrm{AP}})\}$ packets are sent by full-duplex communication in addition to that in the $\tau_{\mathrm{AP}} = 0$ case.

Therefore, the band occupancy for $\frac{\beta_{\mathrm{AP}}(\beta)}{b_{\mathrm{AP}}}\{1 - \exp(-\lambda_{\mathrm{UT}}\tau_{\mathrm{AP}})\}$ packets is reduced at least.

Then, the band occupancy in practical full-duplex communication with buffering time is satisfied with

$$\begin{aligned} \tilde{\beta} &\leq \beta - \frac{\beta_{\mathrm{AP}}(\beta)}{b}\{1 - \exp(-\lambda_{\mathrm{UT}}\tau_{\mathrm{AP}})\}b \\ &\leq \beta_{\max} - \beta_{\mathrm{AP}}(\beta_{\max})\{1 - \exp(-\lambda_{\mathrm{UT}}\tau_{\mathrm{AP}})\}. \end{aligned} \tag{8}$$

In addition, in the case where user terminal does not set the buffering time ($\tau_{\mathrm{UT}} = 0$), when the access point does not set the buffering time ($\tau_{\mathrm{AP}} = 0$), the band occupancy in practical full-duplex communication with buffering time is also satisfied with

$$\tilde{\beta} \le \beta_{\max} - \beta_{\mathrm{UT}}(\beta_{\max})\{1 - \exp(-\lambda_{\mathrm{AP}}\tau_{\mathrm{UT}})\}. \tag{9}$$

■

Now, we consider the lower bound.

**Theorem 2 (A Lower Bound on Band Occupancy in Practical Full-duplex Communication with Buffering Time):** *The band occupancy of practical full-duplex communication with buffering time is bounded as:*

$$\begin{aligned}
&\tilde{\beta} > \beta_{\min} - \beta_{\mathrm{AP}}(\beta_{\min})(1 - e^{-\lambda_{\mathrm{UT}}\tau_{\mathrm{AP}}}) - \beta_{\mathrm{UT}}(\beta_{\min})(1 - e^{-\lambda_{\mathrm{AP}}\tau_{\mathrm{UT}}}),\\
&\tilde{\beta} > \rho_{\mathrm{AP}},\\
&\tilde{\beta} > \rho_{\mathrm{UT}}.
\end{aligned}$$

*Proof:* When the buffering time is zero for access point and user terminal ($\tau_{\mathrm{AP}} = \tau_{\mathrm{UT}} = 0$), $\frac{\beta_{\mathrm{AP}}(\beta)}{b}$ packets are transmitted by half duplex communication from access point, and $\frac{\beta_{\mathrm{UT}}(\beta)}{b}$ packets are transmitted by half-duplex communication from user terminal. When we assume that access point and user terminal attach the buffering time to all of these half-duplex packets, $\frac{\beta_{\mathrm{AP}}(\beta)}{b}\{1 - \exp(-\lambda_{\mathrm{UT}}\tau_{\mathrm{AP}})\} + \frac{\beta_{\mathrm{UT}}(\beta)}{b}\{1 - \exp(-\lambda_{\mathrm{AP}}\tau_{\mathrm{UT}})\}$, the packets change into full-duplex transmitted packets. However, this is not practical. For example, if access point attaches the buffering time to $\frac{\beta_{\mathrm{AP}}(\beta)}{b}$ originally half-duplex packets, then only $\frac{\beta_{\mathrm{UT}}(\beta)}{b} - \frac{\beta_{\mathrm{AP}}(\beta)}{b}\{1 - \exp(-\lambda_{\mathrm{UT}}\tau_{\mathrm{AP}})\}$ half-duplex packets remain at user terminal. Therefore, the band occupancy in practical full-duplex communication with buffering time is

$$\tilde{\beta} > \beta_{\min} - \beta_{\mathrm{AP}}\left(\beta_{\min}\right)\{1 - \exp(-\lambda_{\mathrm{UT}}\tau_{\mathrm{AP}})\} - \beta_{\mathrm{UT}}(\beta_{\min})\{1 - \exp(-\lambda_{\mathrm{AP}}\tau_{\mathrm{UT}})\}. \tag{10}$$

In addition, each traffic intensity must be less than the band occupancy to ensure system stability, $\rho_{\mathrm{AP}} < \tilde{\beta}$, $\rho_{\mathrm{UT}} < \tilde{\beta}$. ■

*Remark 1:* The results of Theorems 1 and 2 include the result of Lemma 3 when the buffering time is zero for access point and user terminal ($\tau_{\mathrm{AP}} = \tau_{\mathrm{UT}} = 0$).

## IV. Mean Waiting Time

In this section, we determine the mean waiting time for each of the four system models. We consider a packet that arrives at the access point (or the user terminal), which we refer to as an arriving packet. The waiting time for the arriving packet depends on the packets that are already in the access point and user terminal when the packet arrives. In particular, a packet is being transmitted (which we refer to as a transmitting packet) and there are some packets in the access point and user terminal queues.

First, we consider the transmitting packet when the arriving packet arrives. The arriving packet needs to wait until the end of the packet's transmission. This is the residual service time for packet transmission, but not the total time length for the transmitting packet. Next, we show the mean residual service time under Poisson arrival traffic.

**Lemma 5 (Mean Residual Service Time for a Deterministic Service):** *The average mean residual service time for a deterministic service under Poisson arrival is $b/2$ for a deterministic service with b. In general, the mean residual service time of a service under Poisson arrival is* $\mathrm{E}[\textit{residual service time}] = \frac{\mathrm{E}[S^2]}{2\mathrm{E}[S]}$, *where $S$ is the time length of the service and* $\mathrm{E}[S]$ *is the average service length [21].*

Thus, we can obtain the mean waiting time in half-duplex communication.

**Lemma 6 (Mean Waiting Time under Half-duplex Communication):** *The mean waiting time for access point under half-duplex communication is* $\mathrm{E}[W_{\mathrm{AP,HD}}] = \frac{\beta_{\mathrm{HD}} b_{\mathrm{HD}}}{2(1-\beta_{\mathrm{HD}})} + b$, *where* $\lambda_{\mathrm{HD}} = \lambda_{\mathrm{AP}} + \lambda_{\mathrm{UT}}$ *and* $b_{\mathrm{HD}} = \frac{\beta \mathrm{HD}}{\lambda_{\mathrm{HD}}}$. *The mean waiting time for user terminal under half-duplex communication is* $\mathrm{E}[W_{\mathrm{UT,HD}}] = \frac{\beta_{\mathrm{HD}} b_{\mathrm{HD}}}{2(1-\beta_{\mathrm{HD}})} + b$.

*Proof:* See Appendix A. ∎

In addition, the mean waiting time can be calculated for ideal full-duplex communication.

**Lemma 7 (Mean Waiting Time under Ideal Full-duplex Communication):** *The mean waiting time for access point under ideal full-duplex communication is* $\mathrm{E}[W_{\mathrm{AP,IFD}}] = \frac{\rho_{\mathrm{AP}} b}{2(1-\rho_{\mathrm{AP}})} + b$. *Similarly, the mean waiting time for user terminal under ideal full-duplex communication is* $\mathrm{E}[W_{\mathrm{UT,IFD}}] = \frac{\rho_{\mathrm{UT}} b}{2(1-\rho_{\mathrm{UT}})} + b$.

*Proof:* See Appendix B. ∎

Thus, we can determine the mean waiting time for practical full-duplex communication without and with buffering time. Before determining the mean waiting time in practical full-duplex communication, we show the mean *queue* waiting time in practical full-duplex communication, which includes that with and without buffering time. Note that the mean queue waiting time in practical full-duplex communication can be adapted to that without buffering time when the buffering time is zero ($\tau_{\mathrm{AP}} = \tau_{\mathrm{UT}} = 0$).

**Lemma 8 (Mean Queue Waiting Time in Practical Full-duplex Communication):** *The mean queue waiting time for access point under practical full-duplex communication is given by (11) and that for user terminal is given by (12), where $\tilde{\beta}_{\mathrm{AP}}\{= \beta_{\mathrm{AP}}(\tilde{\beta})\}$, $\tilde{\beta}_{\mathrm{FD}}\{= \beta_{\mathrm{FD}}(\tilde{\beta})\}$, and $\tilde{\beta}_{\mathrm{UT}}\{= \beta_{\mathrm{UT}}(\tilde{\beta})\}$ are the band occupancies in half-duplex transmission from access point, full-duplex communication, and half-duplex transmission from access point, respectively. In addition, $\tilde{\gamma}_{\tau,\mathrm{AP}}$and $\tilde{\gamma}_{\tau,\mathrm{UT}}$ are the fractions of the halting duration for the buffering time of access point and user terminal, respectively.*

$$\mathrm{E}[W_{\mathrm{AP}}, q] = \frac{1}{2\left\{1 - \left(\rho_{\mathrm{AP}} + \tilde{\gamma}_{\tau,\mathrm{AP}}\right)\right\}} \left(\tilde{\beta} b + \tilde{\gamma}_{\tau,\mathrm{UT}}\tau_{\mathrm{UT}} + \tilde{\gamma}_{\tau,\mathrm{AP}}\tau_{\mathrm{AP}} + 2\tilde{\gamma}_{\tau,\mathrm{AP}} b\right) \tag{11}$$

$$\mathrm{E}[W_{\mathrm{UT}}, q] = \frac{1}{2\left\{1 - \left(\rho_{\mathrm{UT}} + \tilde{\gamma}_{\tau,\mathrm{UT}}\right)\right\}} \left(\tilde{\beta} b + \tilde{\gamma}_{\tau,\mathrm{UT}}\tau_{\mathrm{UT}} + \tilde{\gamma}_{\tau,\mathrm{AP}}\tau_{\mathrm{AP}} + 2\tilde{\gamma}_{\tau,\mathrm{UT}} b\right) \tag{12}$$

*Proof:* Consider a packet that arrives in the access point queue. The delay for the packet is determined by the other packets in the system when the packet arrives. Thus, the packet (the serving packet) is served when the packet arrives on the access point (the arrived packet). The time remaining for the serving packet affects the arrived packet.

Now, we consider a packet that arrives in the queue. Each packet in the queue affects the delay of the arrived packet. The access point queue length when the packet arrives is $\mathrm{E}[L_{\mathrm{AP},q}]$ because of the "PASTA" rule [20]. Thus, the delay of the arrived packet is $\hat{b}_{\mathrm{AP}}\mathrm{E}[L_{\mathrm{AP},q}]$ due to the packets in the access point queue, where $\hat{b}_{\mathrm{AP}}$ is the average duration between the start and end of the service. Note that the average duration between the start and end of the service ($\hat{b}_{\mathrm{AP}}$) is not equal to $b$ because the average ($\hat{b}_{\mathrm{AP}}$) includes the buffering time ($\tau_{\mathrm{AP}}$).

Therefore, the average access point queue waiting time for the arrived packet is obtained by (13) because the PASTA property indicates that each probability for the arrived packet is the same as each fraction of time. Using $\mathrm{E}[L_{\mathrm{AP},q}] = \lambda_{\mathrm{AP}}\mathrm{E}[W_{\mathrm{AP},q}]$ [22], the average access point queue waiting time can be reformulated as (14).

$$\begin{aligned}\mathrm{E}[W_{\mathrm{AP},q}] &= \tilde{\beta}_{\mathrm{UT}}\frac{b}{2} + \tilde{\beta}_{\mathrm{FD}}\frac{b}{2} + \tilde{\beta}_{\mathrm{AP}}\frac{b}{2} + \tilde{\gamma}_{\tau,\mathrm{UT}}\frac{\tau_{\mathrm{UT}}}{2} + \tilde{\gamma}_{\tau,\mathrm{AP}}\left(\frac{\tau_{\mathrm{AP}}}{2} + b\right) + \hat{b}_{\mathrm{AP}}\mathrm{E}[L_{\mathrm{AP},q}] \qquad (13)\\ &= \frac{1}{2\left(1 - \hat{b}_{\mathrm{AP}}\lambda_{\mathrm{AP}}\right)} \left(\tilde{\beta} b + \tilde{\gamma}_{\tau,\mathrm{UT}}\tau_{\mathrm{UT}} + \tilde{\gamma}_{\tau,\mathrm{AP}}\tau_{\mathrm{AP}} + 2\tilde{\gamma}_{\tau,\mathrm{AP}} b\right) \qquad (14)\end{aligned}$$

The average access point duration between the start and end of the service ($\hat{b}_{\mathrm{AP}}$) is the total service duration divided by the number of all the packets that have arrived at access point. Thus, the average duration between the start and end of the service is

$$\hat{b}_{\mathrm{AP}} = \frac{1}{\lambda_{\mathrm{AP}}}(\tilde{\beta}_{\mathrm{AP}} + \tilde{\beta}_{\mathrm{FD}} + \tilde{\gamma}_{\tau,\mathrm{AP}}). \tag{15}$$

Then, by using (15) to eliminate $\hat{b}_{\mathrm{AP}}$, we can obtain the average access point queue waiting time (11). In the same manner as the average access point queue waiting time, the average user terminal queue waiting time can be obtained as (12). ■

**Theorem 3 (An Upper Bound on the Mean Waiting Time under Practical Full-duplex Communication):** *An upper bound on the mean waiting time for packets that arrives at access point is shown by (16),*

$$\mathrm{E}[W_{\mathrm{AP}}] < \frac{1}{2\left\{1-(\rho_{\mathrm{AP}}+\tilde{\gamma}_{\tau,\mathrm{AP,max}})\right\}}\left(\tilde{\beta}_{\max}b + D_{\max} + 2\tilde{\gamma}_{\tau,\mathrm{AP,max}}b\right) + \frac{1}{\lambda_{\mathrm{AP}}}(\rho_{\mathrm{AP}} + \tilde{\gamma}_{\tau,\mathrm{AP,max}}) \tag{16}$$

*where $\tilde{\beta}_{\max}$ is the upper bound on band occupancy in practical full-duplex communication (see Theorem 1) and $\tilde{\beta}_{\min}$ is the lower bound on band occupancy in practical full-duplex communication (see Theorem 2), which are defined by (17) and (18), respectively, $D_{\max}$ represents the upper bound on $\tilde{\gamma}_{\tau,\mathrm{UT}}\tau_{\mathrm{UT}}+\tilde{\gamma}_{\tau,\mathrm{AP}}\tau_{\mathrm{AP}}(=D)$, as shown by (19) and $\tilde{\gamma}_{\tau,\mathrm{AP,max}} = \min\left\{\beta_{\mathrm{AP}}(\beta_{\max})\frac{\tau_{\mathrm{AP}}}{b}, 1-\tilde{\beta}_{\min}\right\}$.*

$$\tilde{\beta}_{\max} = \begin{cases} \min\left\{1-(1-\rho_{\mathrm{UT}})\left(1-e^{-\lambda_{\mathrm{UT}}\tau_{\mathrm{AP}}}\right), 1-(1-\rho_{\mathrm{UT}})\left(1-e^{-\lambda_{\mathrm{AP}}\tau_{\mathrm{UT}}}\right)\right\}, & \text{for } \rho_{\mathrm{AP}}+\rho_{\mathrm{UT}} \geq 1 \\ \rho_{\mathrm{AP}}+\rho_{\mathrm{UT}} - \max\left\{\rho_{\mathrm{AP}}\left(1-e^{-\lambda_{\mathrm{UT}}\tau_{\mathrm{AP}}}\right), \rho_{\mathrm{UT}}\left(1-e^{-\lambda_{\mathrm{AP}}\tau_{\mathrm{UT}}}\right)\right\}, & \text{for } \rho_{\mathrm{AP}}+\rho_{\mathrm{UT}} < 1 \end{cases} \tag{17}$$

$$\tilde{\beta}_{\min} = \max\left\{\beta_{\min} - \beta_{\mathrm{AP}}(\beta_{\max})(1-e^{-\lambda_{\mathrm{UT}}\tau_{\mathrm{AP}}}) - \beta_{\mathrm{UT}}(\beta_{\max})(1-e^{-\lambda_{\mathrm{AP}}\tau_{\mathrm{UT}}}), \rho_{\mathrm{AP}}, \rho_{\mathrm{UT}}\right\} \tag{18}$$

$$D_{\max} = \min\left\{\beta_{\mathrm{AP}}(\beta_{\max})\frac{\tau_{\mathrm{AP}}^2}{b} + \beta_{\mathrm{UT}}(\beta_{\max})\frac{\tau_{\mathrm{UT}}^2}{b}, (1-\tilde{\beta}_{\min})\max(\tau_{ap}, \tau_{ut})\right\} \tag{19}$$

*An upper bound on the mean waiting time for user terminal packets is given by (20)*

$$\mathrm{E}[W_{\mathrm{UT}}] < \frac{1}{2\left[1-\{\rho_{\mathrm{UT}}+\tilde{\gamma}_{\tau,\mathrm{UT,max}}\}\right]}\left(\tilde{\beta}_{\max}b + D_{\max} + 2\tilde{\gamma}_{\tau,\mathrm{UT,max}}b\right) + \frac{1}{\lambda_{\mathrm{UT}}}(\rho_{\mathrm{UT}} + \tilde{\gamma}_{\tau,\mathrm{UT,max}}) \tag{20}$$

*where $\tilde{\gamma}_{\tau,\mathrm{UT,max}} = \min\left\{\beta_{\mathrm{UT}}(\beta_{\max})\frac{\tau_{\mathrm{UT}}}{b}, 1-\tilde{\beta}_{\min}\right\}$.*

*Proof:* To obtain the upper bound on the mean waiting time, we calculate the upper bound of the band occupancy in half-duplex transmission from user terminal ($\tilde{\beta}_{\mathrm{UT}}$), that in full-duplex transmission ($\tilde{\beta}_{\mathrm{FD}}$), and that in half-duplex transmission from access point ($\tilde{\beta}_{\mathrm{AP}}$), as well as the fractions of the halting duration for the buffering time of user terminal ($\tilde{\gamma}_{\tau,\mathrm{UT}}$) and that of access point ($\tilde{\gamma}_{\tau,\mathrm{UT}}$).

First, we obtain the band occupancy in half-duplex transmission and that in full-duplex transmission from Lemma 4. The sum of the band occupancy in half-duplex transmission from access point and full-duplex transmission is

$$\tilde{\beta}_{\mathrm{AP}} + \tilde{\beta}_{\mathrm{FD}} = \rho_{\mathrm{AP}}, \tag{21}$$

because the length of full-duplex communication is defined as $b_{\mathrm{FD}} = b_{\mathrm{AP}}$. The band occupancy in half-duplex transmission from user terminal is bounded as

$$\tilde{\beta}_{\mathrm{UT}} < \beta_{\mathrm{UT}}(\tilde{\beta}_{\max}). \tag{22}$$

Second, we consider the fractions of halting duration for the buffering time of access point. The fraction of the halting duration depends on the number of packets transmitted from access point. When access point operates practical full-duplex communication without buffering time, $\frac{\beta_{\mathrm{AP}}(\beta)}{b}$ packets are transmitted as half-duplex from access point. The fractions of the halting duration for the buffering time of access point are bounded as follows:

$$\begin{aligned}\tilde{\gamma}_{\tau,\mathrm{AP}} &< \beta_{\mathrm{AP}}(\beta)\frac{\tau_{\mathrm{AP}}}{b} \\ &< \beta_{\mathrm{AP}}(\beta_{\max})\frac{\tau_{\mathrm{AP}}}{b}(:= \tilde{\gamma}_{\tau,\mathrm{AP},\max}).\end{aligned} \tag{23}$$

In the same manner, the fractions of the halting duration for the buffering time of user terminal are bounded as follows:

$$\tilde{\gamma}_{\tau,\mathrm{UT}} < \beta_{\mathrm{UT}}(\beta_{\max})\frac{\tau_{\mathrm{UT}}}{b}(:= \tilde{\gamma}_{\tau,\mathrm{UT},\max}). \tag{24}$$

In addition, the summed fractions of the halting duration for the buffering times of access point and user terminal are also bounded as follows:

$$\tilde{\gamma}_{\tau,\mathrm{AP}} + \tilde{\gamma}_{\tau,\mathrm{UT}} < 1 - \tilde{\beta} < 1 - \tilde{\beta}_{\min}, \tag{25}$$

because the summed band occupancy and the halting fractions cannot exceed 1.

Thus, we can obtain the upper bound on $\tilde{\gamma}_{\tau,\mathrm{UT}}\tau_{\mathrm{UT}} + \tilde{\gamma}_{\tau,\mathrm{AP}}\tau_{\mathrm{AP}}$ (= $D$). By using (23) and (24) to eliminate $\tilde{\gamma}_{\tau,\mathrm{UT}}$ and $\tilde{\gamma}_{\tau,\mathrm{AP}}$, we can obtain

$$D < \beta_{\mathrm{UT}}(\beta_{\max})\frac{\tau_{\mathrm{UT}}^2}{b} + \beta_{\mathrm{AP}}(\beta_{\max})\frac{\tau_{\mathrm{AP}}^2}{b}. \tag{26}$$

Moreover, $D$ can be bounded in the other direction. By using (25) to eliminate $\tilde{\gamma}_{\tau,\mathrm{UT}} + \tilde{\gamma}_{\tau,\mathrm{AP}}$, $D$ is bounded as follows:

$$\begin{aligned}D &< (\tilde{\gamma}_{\tau,\mathrm{UT}} + \tilde{\gamma}_{\tau,\mathrm{AP}})\max(\tau_{\mathrm{AP}}, \tau_{\mathrm{UT}}) \\ &< (1 - \tilde{\beta}_{\min})\max(\tau_{\mathrm{AP}}, \tau_{\mathrm{UT}}).\end{aligned} \tag{27}$$

We describe these bounds as $D < D_{\max}$.

The relationships between the mean queue waiting time and mean waiting time are

$$\mathrm{E}[W_{\mathrm{AP}}] = \mathrm{E}[W_{\mathrm{AP}}, q] + \hat{b}_{\mathrm{AP}}, \tag{28}$$

$$\mathrm{E}[W_{\mathrm{UT}}] = \mathrm{E}[W_{\mathrm{UT}}, q] + \hat{b}_{\mathrm{UT}}, \tag{29}$$

where $\hat{b}_{\mathrm{UT}} = \frac{1}{\lambda_{\mathrm{UT}}}(\tilde{\beta}_{\mathrm{UT}} + \tilde{\beta}_{\mathrm{FD}} + \tilde{\gamma}_{\tau,\mathrm{UT}})$. Thus, we can express $\mathrm{E}[W_{\mathrm{AP}}]$ as shown in (31),by combining (28), (11) in Lemma 8, (21), (22), (23), (26), and (27). We obtain the upper bound on the mean waiting time for access point packets as shown by (16). In the same manner, we can express $\mathrm{E}[W_{\mathrm{UT}}]$ as shown in (33) by combining (29), (11) in Lemma 8, (21), (22), (24), (26), and (27). We can obtain the upper bound on the mean waiting time for user terminal packets as shown by (20).

$$\begin{aligned}
\mathrm{E}[W_{\mathrm{AP}}] &= \frac{1}{2\left\{1-\left(\rho_{\mathrm{AP}}+\tilde{\gamma}_{\tau,\mathrm{AP}}\right)\right\}}\left(\tilde{\beta}b + \tilde{\gamma}_{\tau,\mathrm{UT}}\tau_{\mathrm{UT}} + \tilde{\gamma}_{\tau,\mathrm{AP}}\tau_{\mathrm{AP}} + 2\tilde{\gamma}_{\tau,\mathrm{AP}}b\right) + \frac{1}{\lambda_{\mathrm{AP}}}\left(\rho_{\mathrm{AP}}+\tilde{\gamma}_{\tau,\mathrm{AP}}\right) && (30)\\
&< \frac{1}{2\left\{1-\left(\rho_{\mathrm{AP}}+\tilde{\gamma}_{\tau,\mathrm{AP}}\right)\right\}}\left(\tilde{\beta}_{\max}b + \tilde{\gamma}_{\tau,\mathrm{UT}}\tau_{\mathrm{UT}} + \tilde{\gamma}_{\tau,\mathrm{AP}}\tau_{\mathrm{AP}} + 2\tilde{\gamma}_{\tau,\mathrm{AP}}b\right) + \frac{1}{\lambda_{\mathrm{AP}}}\left(\rho_{\mathrm{AP}}+\tilde{\gamma}_{\tau,\mathrm{AP}}\right)\\
&< \frac{1}{2\left\{1-\left(\rho_{\mathrm{AP}}+\tilde{\gamma}_{\tau,\mathrm{AP,max}}\right)\right\}}\left(\tilde{\beta}_{\max}b + \tilde{\gamma}_{\tau,\mathrm{UT}}\tau_{\mathrm{UT}} + \tilde{\gamma}_{\tau,\mathrm{AP}}\tau_{\mathrm{AP}} + 2\tilde{\gamma}_{\tau,\mathrm{AP,max}}b\right) + \frac{1}{\lambda_{\mathrm{AP}}}\left(\rho_{\mathrm{AP}}+\tilde{\gamma}_{\tau,\mathrm{AP,max}}\right)\\
&< \frac{1}{2\left\{1-\left(\rho_{\mathrm{AP}}+\tilde{\gamma}_{\tau,\mathrm{AP,max}}\right)\right\}}\left(\tilde{\beta}_{\max}b + D_{\max} + 2\tilde{\gamma}_{\tau,\mathrm{AP,max}}b\right) + \frac{1}{\lambda_{\mathrm{AP}}}\left(\rho_{\mathrm{AP}}+\tilde{\gamma}_{\tau,\mathrm{AP,max}}\right) && (31)
\end{aligned}$$

$$\begin{aligned}
\mathrm{E}[W_{\mathrm{UT}}] &= \frac{1}{2\left\{1-\left(\rho_{\mathrm{UT}}+\tilde{\gamma}_{\tau,\mathrm{UT}}\right)\right\}}\left\{\tilde{\beta}b + \tilde{\gamma}_{\tau,\mathrm{UT}}\left(\tau_{\mathrm{UT}}+2b\right) + \tilde{\gamma}_{\tau,\mathrm{AP}}\tau_{\mathrm{AP}}\right\} + \frac{1}{\lambda_{\mathrm{UT}}}\left(\rho_{\mathrm{UT}}+\tilde{\gamma}_{\tau,\mathrm{UT}}\right) && (32)\\
&< \frac{1}{2\left\{1-\left(\rho_{\mathrm{UT}}+\tilde{\gamma}_{\tau,\mathrm{UT,max}}\right)\right\}}\left(\tilde{\beta}_{\max}b + D_{\max} + 2\tilde{\gamma}_{\tau,\mathrm{UT,max}}b\right) + \frac{1}{\lambda_{\mathrm{AP}}}\left(\rho_{\mathrm{AP}}+\tilde{\gamma}_{\tau,\mathrm{UT,max}}\right) && (33)
\end{aligned}$$

■

Next, we obtain a lower bound on the mean waiting time for practical full-duplex communication.

**Theorem 4 (A Lower Bound on the Mean Waiting Time for Practical Full-duplex Communication):** *A lower bound on the mean waiting time for access point under practical full-duplex communication is given by (34), where* $\tilde{\gamma}_{\tau,\mathrm{AP,min}}$ *is defined as* $\tilde{\gamma}_{\tau,\mathrm{AP,min}} = \min\left\{\frac{\beta_{\mathrm{AP}}(\tilde{\beta}_{\min})}{b}\tau_{\mathrm{AP}}, 1-\tilde{\beta}_{\max}\right\}$.

$$\mathrm{E}[W_{\mathrm{AP}}] > \frac{\tilde{\beta}_{\min}b + \tilde{\gamma}_{\tau,\mathrm{UT,min}}\tau_{\mathrm{UT}} + \tilde{\gamma}_{\tau,\mathrm{AP,min}}\left(\tau_{\mathrm{AP}}+2b\right)}{2\left\{1-\left(\rho_{\mathrm{AP}}+\tilde{\gamma}_{\tau,\mathrm{AP,min}}\right)\right\}} + \frac{\rho_{\mathrm{AP}}+\tilde{\gamma}_{\tau,\mathrm{AP,min}}}{\lambda_{\mathrm{AP}}} \tag{34}$$

*The lower bound on the mean waiting time for user terminal under practical full-duplex communication is given by (35), where* $\tilde{\gamma}_{\tau,\mathrm{UT,min}} = \min\left\{\frac{\beta_{\mathrm{UT}}(\tilde{\beta}_{\min})}{b}\tau_{\mathrm{UT}}, 1-\tilde{\beta}_{\max}\right\}$.

$$\mathrm{E}[W_{\mathrm{UT}}] > \frac{\tilde{\beta}_{\min}b + \tilde{\gamma}_{\tau,\mathrm{UT,min}}\left(\tau_{\mathrm{UT}}+2b_{\mathrm{UT}}\right) + \tilde{\gamma}_{\tau,\mathrm{AP,min}}\tau_{\mathrm{AP}}}{2\left\{1-\left(\rho_{\mathrm{UT}}+\tilde{\gamma}_{\tau,\mathrm{UT,min}}\right)\right\}} + \frac{\rho_{\mathrm{UT}}+\tilde{\gamma}_{\tau,\mathrm{UT,min}}}{\lambda_{\mathrm{UT}}} \tag{35}$$

*Proof:* We determine the lower bound of $\tilde{\beta}_{\mathrm{UT}}$, $\tilde{\beta}_{\mathrm{UT}} + \tilde{\beta}_{\mathrm{FD}}$, $\tilde{\gamma}_{\tau,\mathrm{AP}}$, and $\tilde{\gamma}_{\tau,\mathrm{UT}}$, before obtaining the lower bound on the mean waiting time for practical full-duplex communication.

The band occupancy in half-duplex transmission from user terminal ($\tilde{\beta}_{\mathrm{UT}}$) is bounded as follows:

$$\tilde{\beta}_{\mathrm{UT}} = \beta_{\mathrm{UT}}(\tilde{\beta}) > \beta_{\mathrm{UT}}(\tilde{\beta}_{\min}). \tag{36}$$

We consider the fractions of the halting duration for the buffering time of access point. The fraction of the halting duration depends on the number of packets transmitted from access point. When the buffering times for access point and user terminal are $\tau_{\mathrm{AP}}$ and $\tau_{\mathrm{UT}}$, respectively, access point sends $\frac{\beta_{\mathrm{AP}}(\tilde{\beta})}{b}$ packets by half-duplex communication. Thus, at least $\frac{\beta_{\mathrm{AP}}(\tilde{\beta}_{\min})}{b} (< \frac{\beta_{\mathrm{AP}}(\tilde{\beta})}{b})$ packets are sent by half-duplex communication from access point. These packets have the buffering time attached. Therefore, $\tilde{\gamma}_{\tau,\mathrm{AP}}$ is bounded as

$$\tilde{\gamma}_{\tau,\mathrm{AP}} \geq \frac{\beta_{\mathrm{AP}}(\tilde{\beta}_{\min})}{b} \tau_{\mathrm{AP}}. \tag{37}$$

In addition, the sum of the band occupancy and the fractions of the halting duration cannot exceed 1, so $\tilde{\gamma}_{\tau,\mathrm{AP}}$ is also bounded as follows:

$$\tilde{\gamma}_{\tau,\mathrm{AP}} \geq 1 - \tilde{\beta}_{\max}. \tag{38}$$

In the same manner as $\tilde{\gamma}_{\tau,\mathrm{AP}}$, $\tilde{\gamma}_{\tau,\mathrm{UT}}$ is bounded as follows:

$$\tilde{\gamma}_{\tau,\mathrm{UT}} \geq \frac{\beta_{\mathrm{UT}}(\tilde{\beta}_{\min})}{b} \tau_{\mathrm{UT}}, \tag{39}$$

$$\tilde{\gamma}_{\tau,\mathrm{UT}} \geq 1 - \tilde{\beta}_{\max}. \tag{40}$$

Then, by combining (30) in Theorem 3, (36), (37), and (38), we can obtain (34). Moreover, by combining (32) in Theorem 3, (36), (39), and (40), we can obtain (35). ■

## V. Performance Evaluation

In this section, we present a performance evaluation of full-duplex communication and the effect of the buffering time based on queueing theory and event-driven simulations. The simulation results for each point represent the average of 10 time trials, where each trial evaluated the average of the last 80,000 packet transmission among 100,000 packet transmissions.

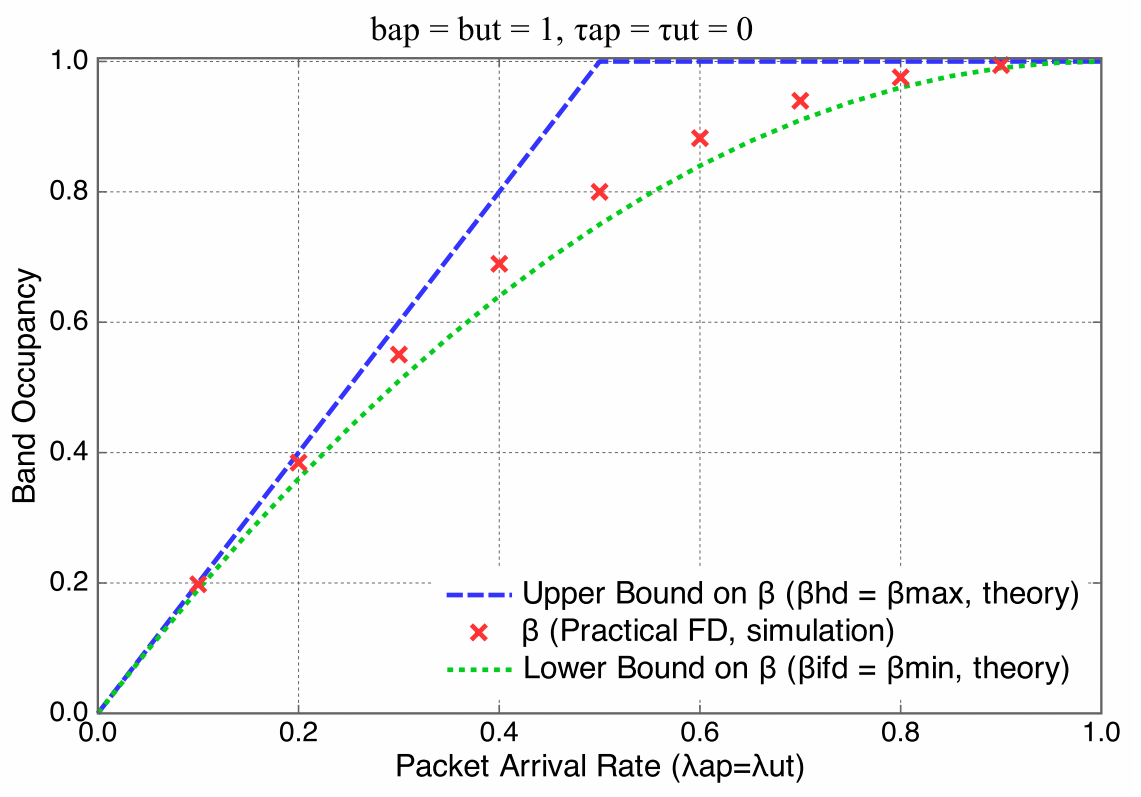

Fig. 8. Packet Arrival Rate vs. Band Occupancy under Symmetrical Traffic

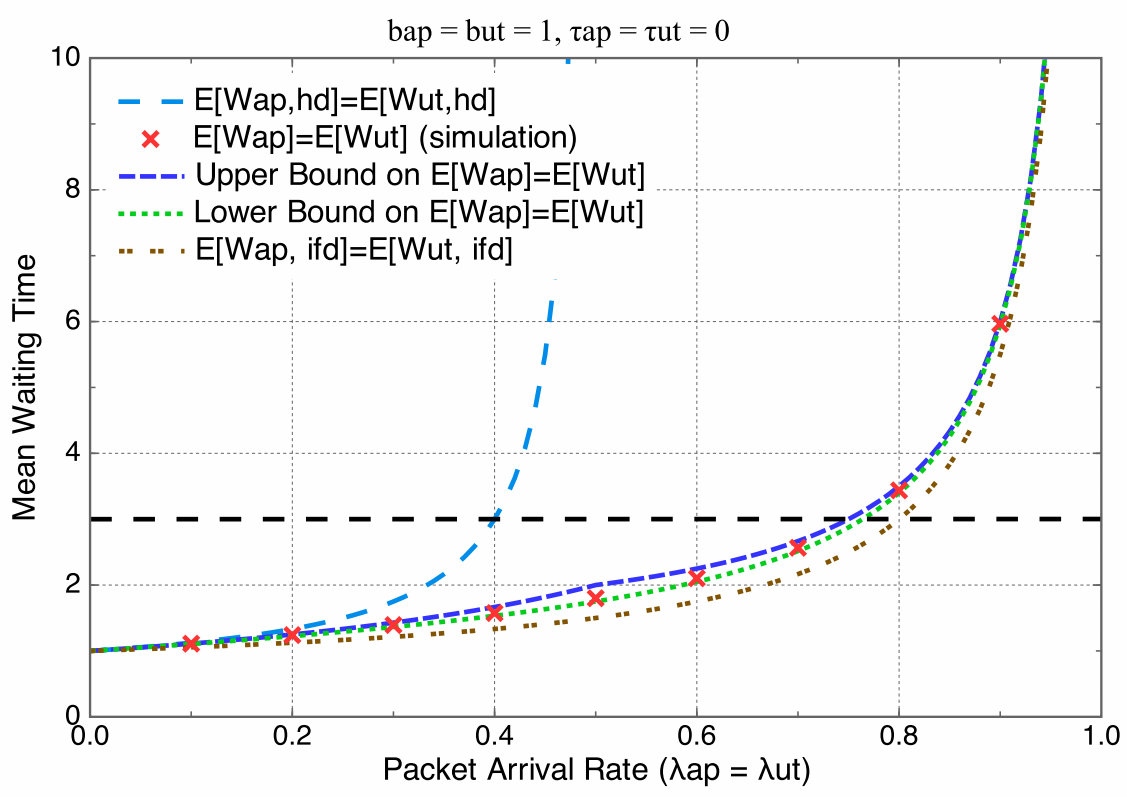

Fig. 9. Packet Arrival Rate vs. Mean Waiting Time under Symmetrical Traffic

### *A. Basic Performance of Practical Full-duplex Communication without Buffering Time*

First, we evaluated the performance under symmetrical traffic to show the basic performance of practical full-duplex communication without buffering time ($\tau_{\text{AP}} = \tau_{\text{AP}} = 0$). The packet arrival rates were the same ($\lambda_{\text{AP}} = \lambda_{\text{UT}}$). We compared half-duplex communication, ideal full-duplex communication, and practical full-duplex communication without buffering time.

Fig. 8 shows the band occupancy when the packet arrival rate changed from 0 to 1. The simulation results in Fig. 8 show that practical full-duplex communication without buffering time only reduced the band occupancy by up to approximately 20 % when the packet arrival rate was 0.5 ($\lambda_{\text{AP}} = \lambda_{\text{UT}} = 0.5$). In addition, the theoretical lower bound suggests that full-duplex communication without buffering time can reduce the band occupancy by up to 25 %. The theoretical upper bound of band occupancy in practical full-duplex communication is equal to the band occupancy in half-duplex communication, and the theoretical lower bound of band occupancy in practical full-duplex communication is equal to the band occupancy in ideal full-duplex communication, as shown in Section III.

Fig. 9 shows the mean waiting time when the packet arrival rate changed from 0 to 1. Theoretically, the mean waiting time in practical full-duplex communication is less than that in half-duplex communication, but larger than that in ideal full-duplex communication (see Appendix C). When the packet arrival rate was 0.5, half-duplex communication was not in a steady state, i.e., the mean waiting time was infinity. However, practical full-duplex communication maintained the mean waiting time at less than 2 when the packet arrival rate was 0.5. Comparing the three system models with the same mean waiting time showed that ideal full-duplex communication doubled the throughput compared with half-duplex communication. However, practical full-duplex communication cannot double the throughput of half-duplex communication because the mean waiting time required for practical full-duplex communication is larger than that for ideal full-duplex communication. For example, the mean waiting time for half-duplex communication was 0.3 when the packet arrival rate was 0.4. In addition, the mean waiting time for ideal full-duplex communication was 0.3 when the packet arrival rate was 0.8. However, the mean waiting time for practical full-duplex communication was 0.3 when the packet arrival rate was approximately 7.5. Practical full-duplex communication improved the throughput by approximately 1.9 times compared with half-duplex communication when the access point and user terminal allowed a mean waiting time of 3.

### *B. Effects of Buffering Time on Band Occupancy*

In the previous subsection, we demonstrated the performance of practical full-duplex communication without buffering time. In this section, we illustrate the effect of the buffering time based on the performance of practical full-duplex communication with buffering time.

To illustrate the basic performance of practical full-duplex communication with buffering time, we evaluated the band occupancy with buffering time. First, we compared practical full-duplex communication with different buffering lengths ($\tau_{\mathrm{UT}} = \tau_{\mathrm{AP}} = 0, 0.5, 1.0$) and the same packet arrival rate for access point and user terminal ($\lambda_{\mathrm{AP}} = \lambda_{\mathrm{UT}}$). The packet length was the same length for access point and user terminal, and for normalization, we set the packet length as a dimensionless quantity of 1 ($b_{\mathrm{AP}} = b_{\mathrm{UT}} = 1$). Fig. 10 shows the packet arrival rate vs. band occupancy with and without buffering time when the packet arrival rate changed from 0 to 1. The simulation results for the band occupancy with different buffering lengths ($\tau_{\rm UT} = \tau_{\rm AP} = 0, 0.5, 1.0$), as well as the theoretical upper and lower bounds of the band occupancy with different buffering lengths ($\tau_{\rm UT} = \tau_{\rm AP} = 0, 0.5$) are shown in the figure.

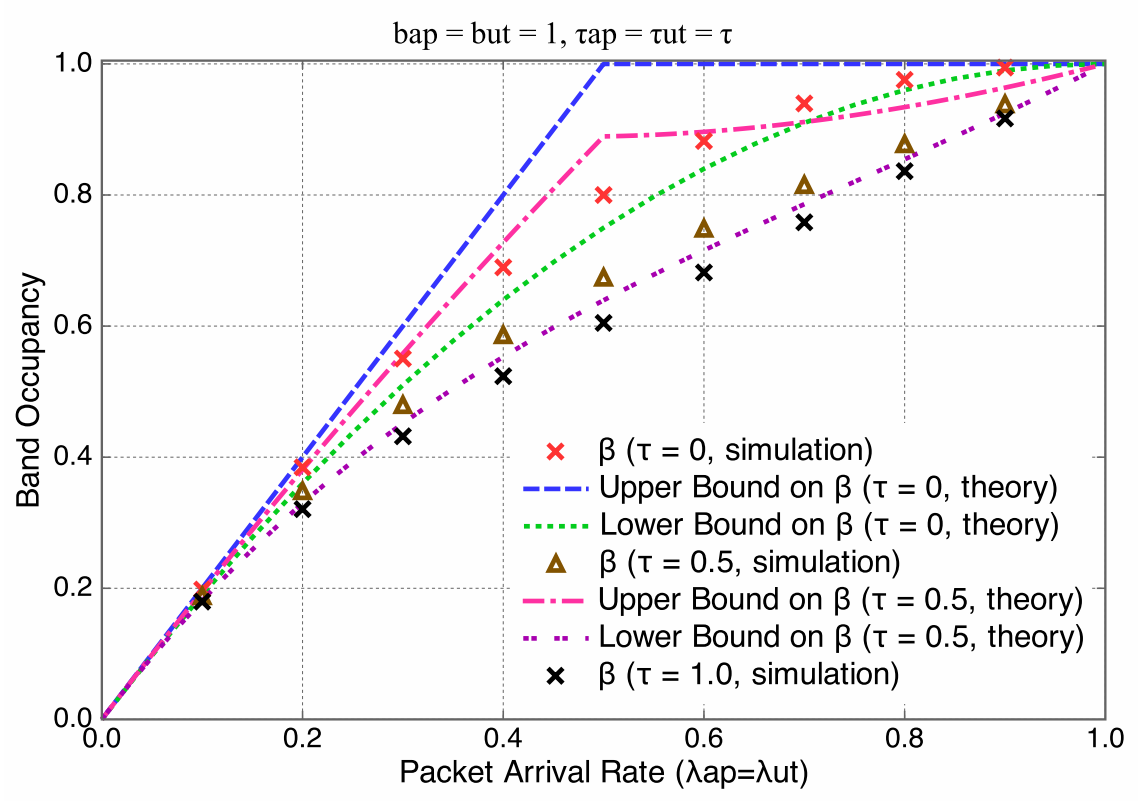

Fig. 10. Packet Arrival Rate vs. Band Occupancy with Buffering Time ($\tau_{\rm UT} = \tau_{\rm AP} = 0, 0.5, 1.0$)

Fig. 10 indicates that the buffering time reduced the band occupancy under any packet arrival rate. For example, when the packet arrival rate was 0.7, the simulation results showed that the buffering time ($\tau_{\rm AP} = \tau_{\rm UT} = 0.5$) reduced the band occupancy by approximately 15 % compared with that without buffering time. The theoretical results also indicate that the upper bound on band occupancy in practical full-duplex with buffering time ($\tau_{\rm AP} = \tau_{\rm UT} = 0.5$) was larger than the lower bound on band occupancy in practical full-duplex without buffering time when the packet arrival rate was 0.7.

However, when the packet arrival rate was low or high, the reduction in band occupancy due to the buffering time was small. When the arrival rate was low, less arrivals occurred during the buffering time. Under high intensity traffic, there was less wastage of the band occupancy in practical full-duplex communication without buffering time. These results suggest that the buffering time should not be attached when the packet arrival rate is low or high.

A comparison of the simulation results with different buffering times of 0.5 and 1.0, as shown in Fig. 10, indicated that the band occupancy did not decrease by a factor of two although the buffering time doubled. Thus, we performed extensive evaluations to determine the effect of the buffering time on band occupancy. We evaluated the band occupancy by changing the buffering time length. A longer buffering time was expected to obtain a greater decrease in the band occupancy. We set the packet arrival rates for access point and user terminal ($\lambda_{\rm AP} = \lambda_{\rm UT}$) as 0.9, 0.5, and 0.2. access point and user terminal had the same packet length, and for the normalization, we set the packet length as a dimensionless quantity of 1 ($b_{\mathrm{AP}} = b_{\mathrm{UT}} = 1$). The traffic intensities in each case ($\rho_{\mathrm{AP}} = \rho_{\mathrm{UT}}$) were derived as 0.9, 0.5, and 0.2.

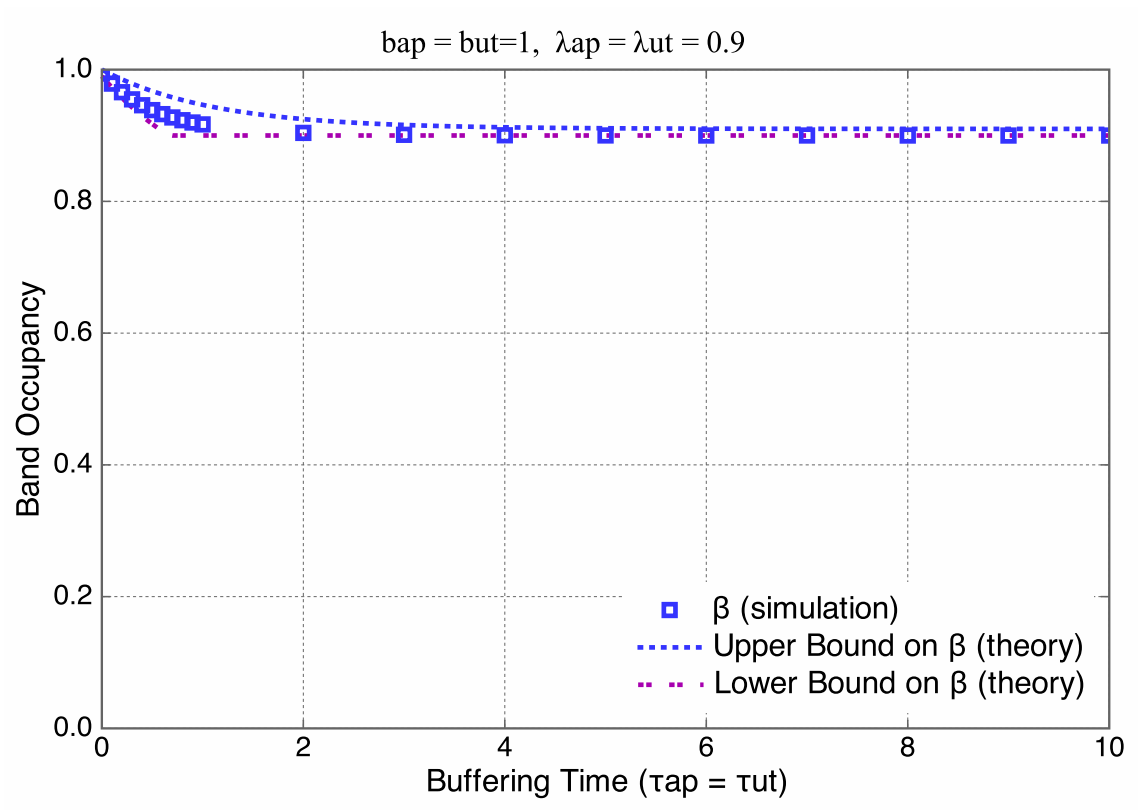

Fig. 11. Buffering Time Length vs. Band Occupancy ($\lambda_{\mathrm{AP}} = \lambda_{\mathrm{UT}} = 0.9$)

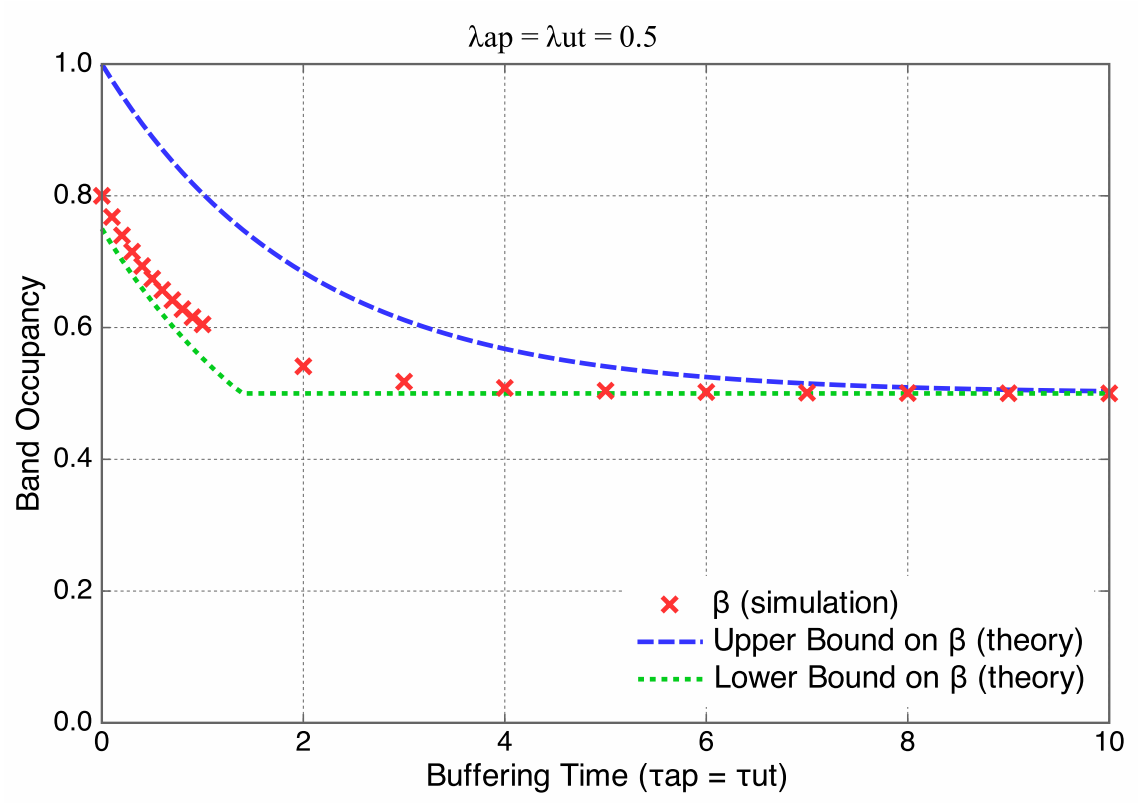

Fig. 12. Buffering Time Length vs. Band Occupancy ($\lambda_{\mathrm{AP}} = \lambda_{\mathrm{UT}} = 0.5$)

Figures 11, 12, and 13 compare the buffering time length with the band occupancy when the packet arrival rate was 0.9, 0.5, and 0.2, respectively. The simulation results and the theoretical upper and lower bounds on the band occupancy are shown in the figures, which demonstrate that the band occupancy decreased exponentially with the buffering time.

Fig. 11 shows that a long buffering time reduced the band occupancy to 0.9, which was the maximum reduction in the band occupancy. Fig. 11 indicates that the buffering time was reduced by a maximum of 50% when the buffering time length was approximately 0.3 and the packet arrival rate was 0.9. Fig. 12 also shows that the buffering time was reduced by a maximum of 50% when the buffering time length was approximately 0.6 and the packet arrival rate was 0.5. Moreover, Fig. 13 shows that the buffering time was decreased by a maximum of 50% when the buffering time length was approximately 1.5 and the packet arrival rate was 0.2.

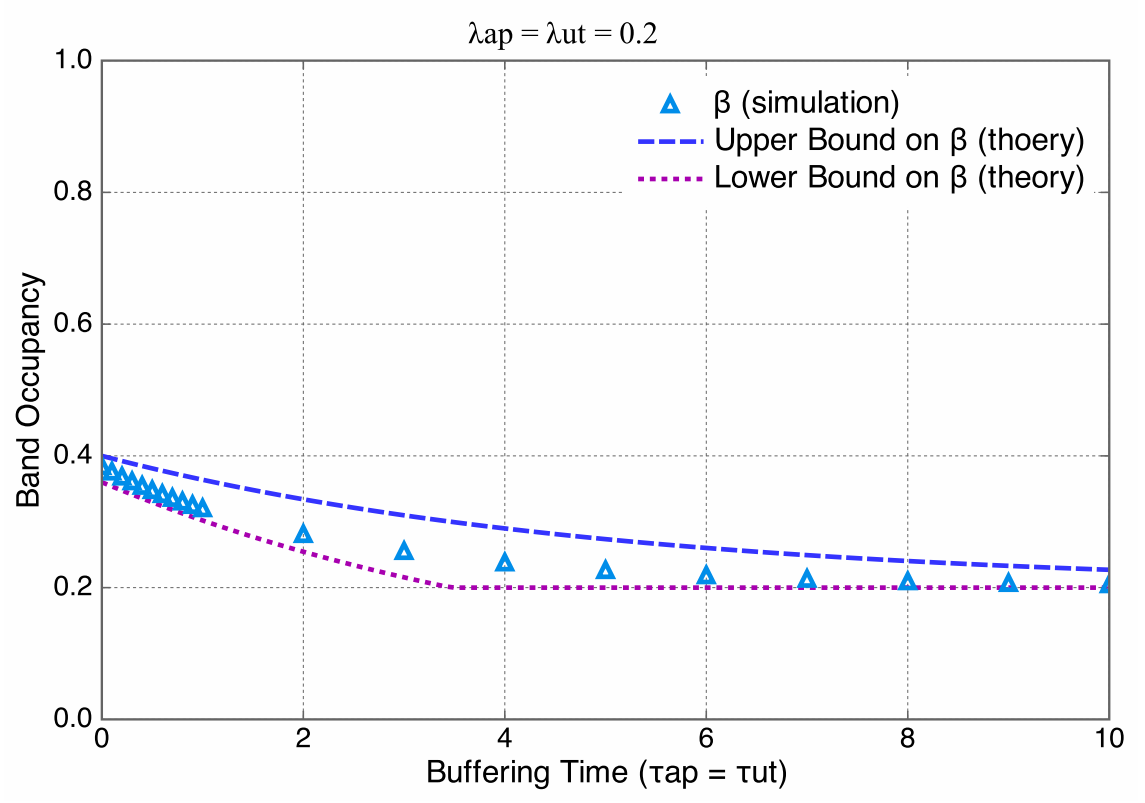

Fig. 13. Buffering Time Length vs. Band Occupancy ($\lambda_{\mathrm{AP}} = \lambda_{\mathrm{UT}} = 0.2$)

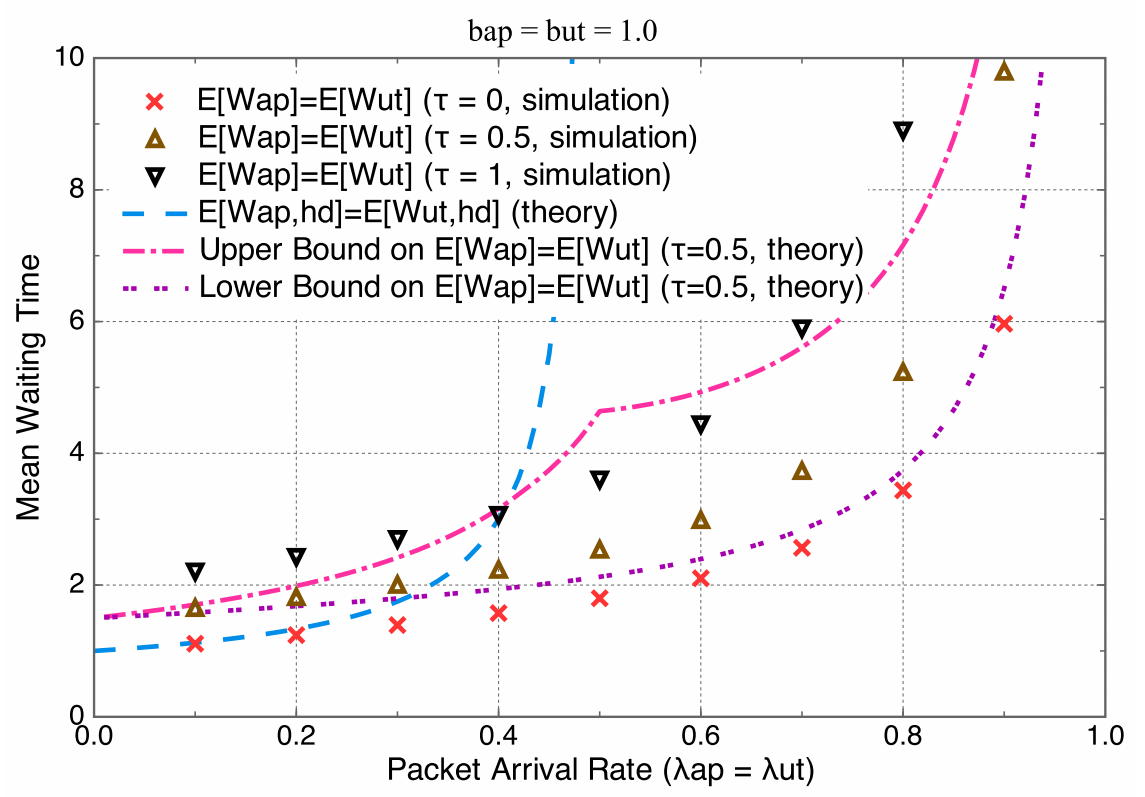

Fig. 14. Packet Arrival Rate vs. Mean Waiting Time

A comparison of Figs. 11, 12, and 13 indicates that the buffering time decreased the band occupancy more efficiently when the traffic intensity was 0.5. This suggests that the buffering time works most efficiently when the traffic is saturated when the sum of the traffic intensity for access point and user terminal is 1 ($\rho_{\mathrm{AP}} + \rho_{\mathrm{UT}} = 1$). Note that full-duplex communication can handle the case where the sum of the traffic intensity exceeds 1.

### *C. Effect of Buffering Time on the Mean Waiting Time*

In the previous subsection, we evaluated the band occupancy with buffering time. To support the performance analysis, we also evaluated the effect of the buffering time on the mean waiting time. We compared practical full-duplex communication with different buffering lengths ($\tau_{\mathrm{UT}} = \tau_{\mathrm{AP}} = 0, 0.5$). The packet arrival rate and packet length were the same length for access point and user terminal ($\lambda_{\mathrm{AP}} = \lambda_{\mathrm{UT}}$, $b_{\mathrm{AP}} = b_{\mathrm{UT}}$). For normalization, we set the packet length as a dimensionless quantity of 1 ($b_{\mathrm{AP}} = b_{\mathrm{UT}} = 1$).

Fig. 14 compares the packet arrival rate with the mean waiting time with buffering time when the packet arrival rate changed from 0 to 1. The results indicate that the buffering time should be changed according to the traffic load and permissible mean waiting time. The permissible mean waiting time depends on the memory size on each node. The theoretical results showed that the mean waiting time in practical full-duplex communication was larger than that in half-duplex communication when the packet arrival rate was less than 0.3.

Thus, for these two examples, the permissible mean waiting times were 3 and 6. We also compared practical full-duplex without buffering time and with buffering time ($\tau_{\mathrm{AP}} = \tau_{\mathrm{UT}} = 1.0$). When the permissible mean waiting time for access point and user terminal was 3, practical full-duplex without buffering time could allow a packet arrival rate of approximately 0.8, whereas practical full-duplex with buffering time allowed a packet arrival rate of approximately 0.4. This suggests that the throughput with buffering time was 0.5x of that without buffering time when the permissible mean waiting time was 3. Note that buffering time reduces band occupancy and also throughput.

When the permissible mean waiting time for access point and user terminal was 6, practical full-duplex without buffering time allowed a packet arrival rate of approximately 0.9, whereas practical full-duplex with buffering time allowed a packet arrival rate of approximately 0.7. Thus, the throughput with buffering time was 0.8x of that without buffering time when the permissible mean waiting time was 6. These two examples suggest that the buffering time should be changed according to the permissible mean waiting time.

We performed extensive evaluations to understand the effects of the buffering time on the mean waiting time in more detail. We evaluated the mean waiting time by changing the buffering time length from 0 to 10. The packet length was the same for access point and user terminal, and for normalization, we set the packet length as a dimensionless quantity of 1 ($b_{\mathrm{AP}} = b_{\mathrm{UT}} = 1$). The traffic intensities in each case ($\rho_{\mathrm{AP}} = \rho_{\mathrm{UT}}$) were derived as 0.9, 0.5, and 0.2.

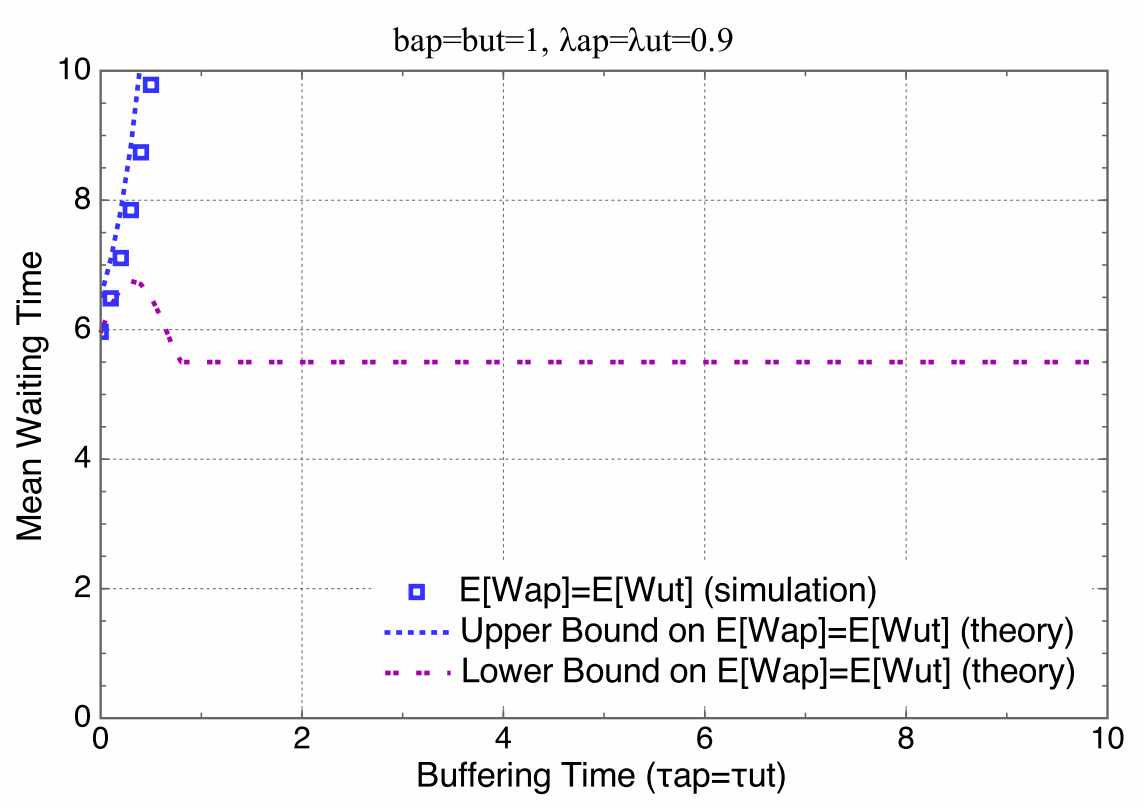

Fig. 15. Buffering Time Length vs. Mean Waiting Time ($\lambda_{\rm AP} = \lambda_{\rm UT} = 0.9$)

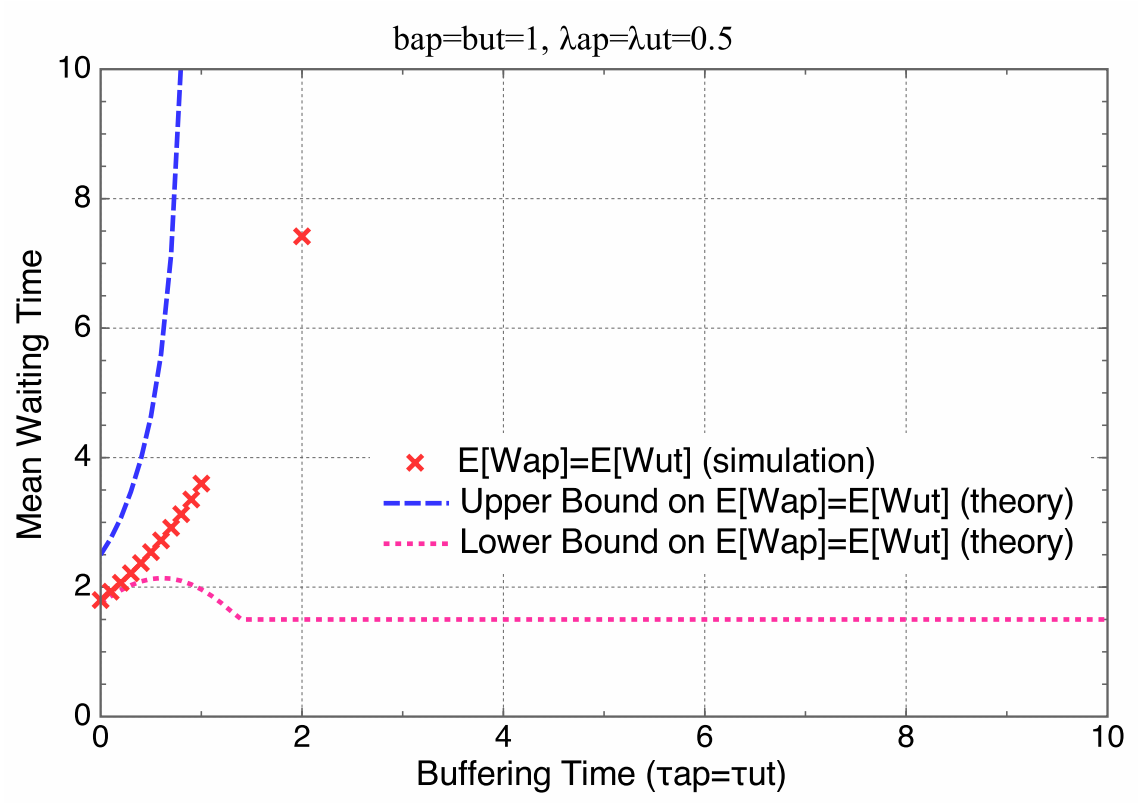

Fig. 16. Buffering Time Length vs. Mean Waiting Time ($\lambda_{\rm AP} = \lambda_{\rm UT} = 0.5$)

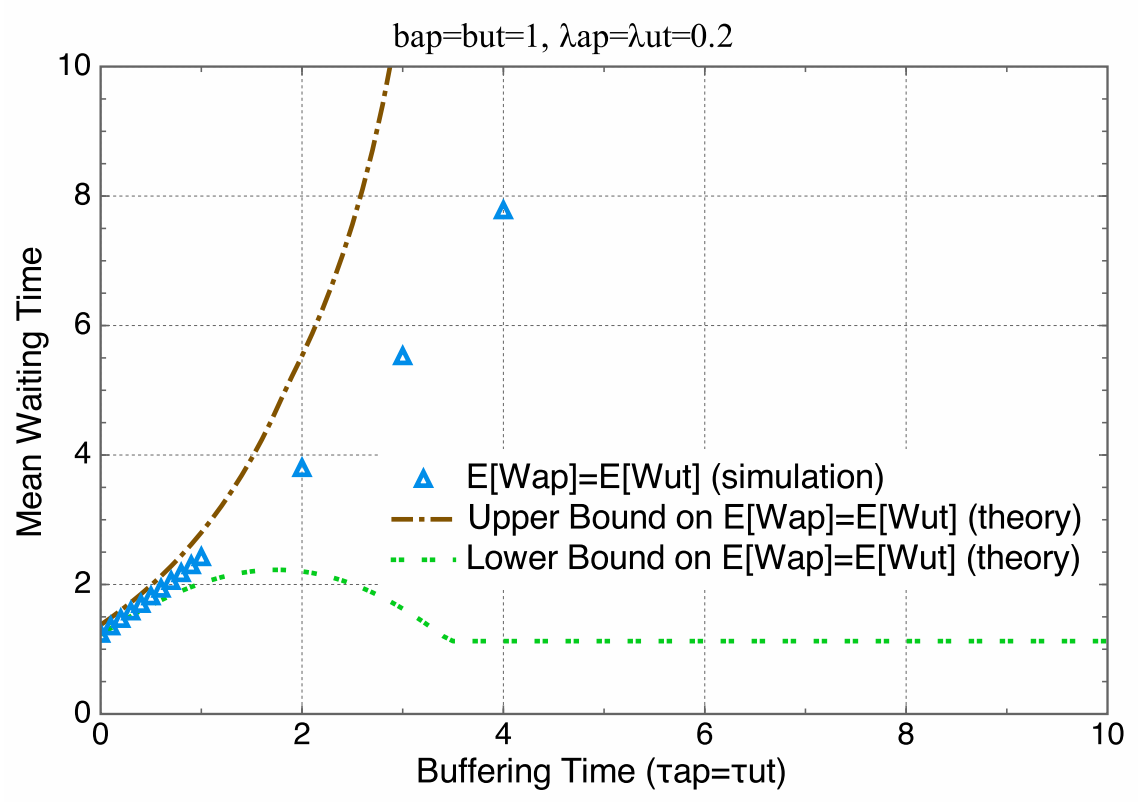

Fig. 17. Buffering Time Length vs. Mean Waiting Time ($\lambda_{\rm AP} = \lambda_{\rm UT} = 0.2$)

Figures 15, 16, and 17 compare the buffering time length with the mean waiting time when the packet arrival rates were 0.9, 0.5, and 0.2, respectively. The simulation results and the theoretical upper and lower bounds on the mean waiting time are shown in the figures.

The simulation results suggest that the mean waiting time increased exponentially with the buffering time length. A comparison of Figures 15, 16, and 17 indicates that the mean waiting time increased more rapidly with a higher packet arrival rate traffic.

Next, we considered the maximum reduction of 50%. In Section V-B, we showed that the buffering time length could achieve a maximum reduction of 50%. According to the results in Section V-B and Figures 15, 16, and 17, we determined the permissible mean waiting time required to achieve a maximum reduction of 50%. The results shown in Figures 11 and 15 suggest that a node memory size that contains approximately eight packets could achieve a maximum reduction of 50% with a packet arrival rate of 0.9. In addition, the results in Figures 12 and 16 suggest that a node memory size that contains approximately three packets could achieve a maximum reduction of 50% with a packet arrival rate of 0.5. Moreover, the results in Figures 13 and 17 show that a node memory size that contains approximately three packets could achieve a maximum reduction of 50% with a packet arrival rate of 0.2. Therefore, a memory size that contains eight packets is sufficient to achieve a maximum reduction in the band occupancy of 50%.

### *D. Effect of Buffering Time under Asymmetrical Traffic*

In the previous section, we illustrated the performance with buffering time under symmetrical traffic. Next, we demonstrate the effect of buffering time under asymmetrical traffic. We evaluated the band occupancy and mean waiting time under asymmetrical traffic. We set the packet arrival rate for access point ($\lambda_{\mathrm{AP}}$) as 0.5 and that for user terminal ($\lambda_{\mathrm{UT}}$) as 0.2 under an assumption of asymmetrical traffic. The packet length was the same for access point and user terminal, and for normalization, we set the packet length as a dimensionless quantity of 1 ($b_{\mathrm{AP}} = b_{\mathrm{UT}} = 1$).

First, we assumed that the buffering time was the same length for access point and user terminal ($\tau_{\mathrm{AP}} = \tau_{\mathrm{UT}}$). Fig. 18 shows the band occupancy under practical full-duplex with buffering time and asymmetric traffic. The simulation results showed that the network system was unstable when the buffering time was longer than approximately 1. When the band occupancy was less than the traffic intensity on each node, the network system was unstable. In addition, the theoretical results showed that the network was unstable when the buffering time was longer than approximately 2.5, i.e., the intersection of the upper and lower bounds.

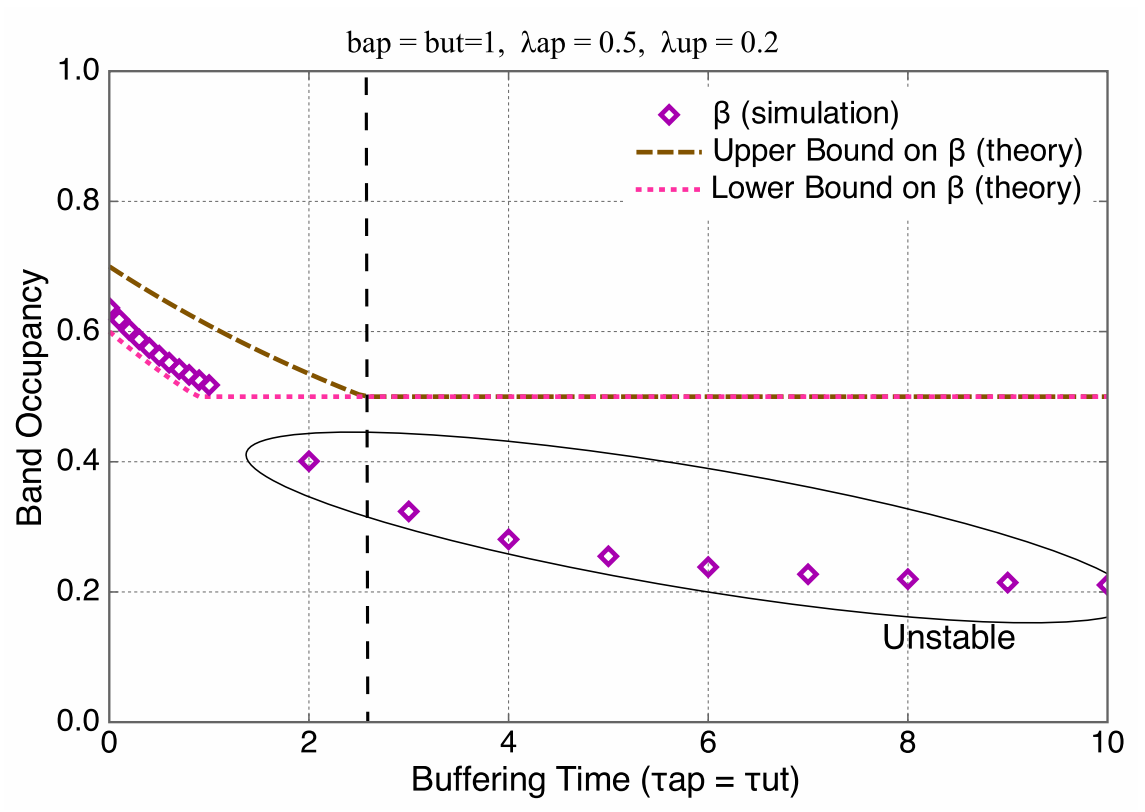

Fig. 18. Buffering Time Length vs. Band Occupancy ($\tau_{\text{AP}} = \tau_{\text{UT}}$)

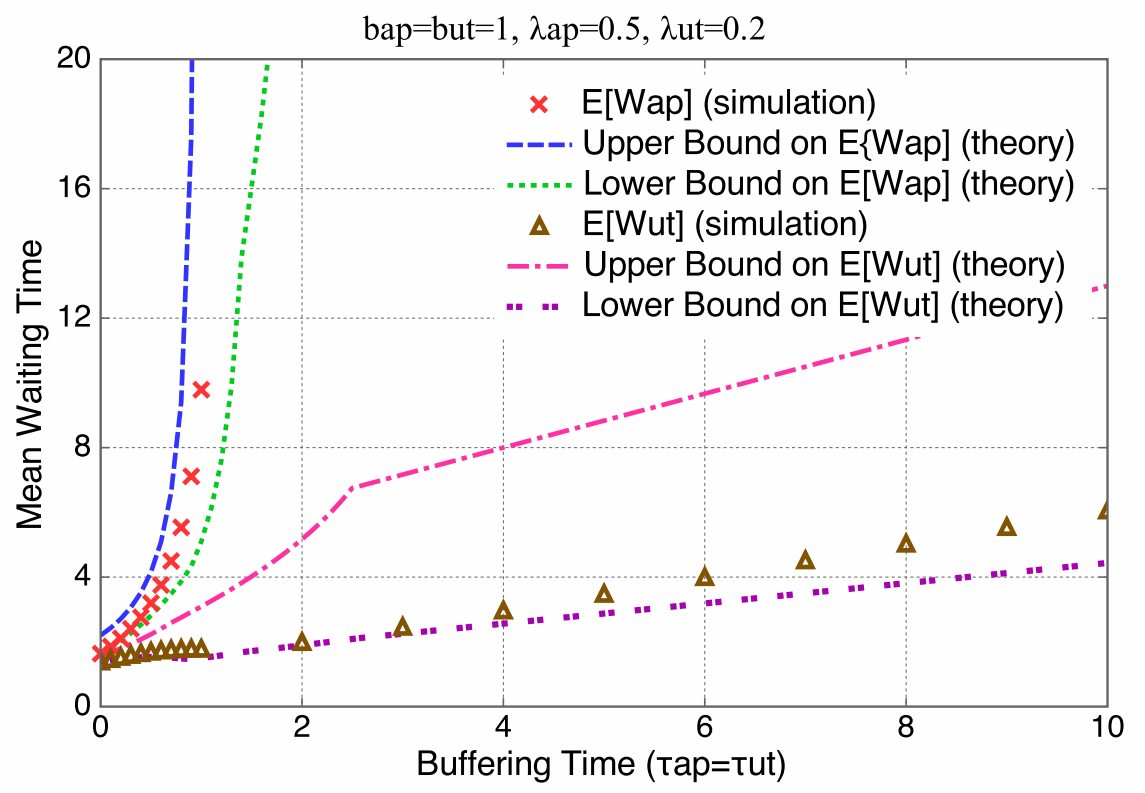

Fig. 19. Buffering Time Length vs. Mean Waiting Time ($\tau_{\text{AP}} = \tau_{\text{UT}}$)

We also evaluated the mean waiting time for practical full-duplex with buffering time. The simulation results and theoretical bounds shown in Fig. 19 indicate that the mean waiting time for access point packets increased exponentially whereas that for user terminal packets increased in a linear manner. Using the same buffering time length for access point and user terminal caused longer delays for access point, which had more packets to send. We suggest that an asymmetrical buffering time could prevent the network system from becoming unstable. We note that user terminals can send data even if the network system becomes unstable.

We assumed that the user terminals set the buffering time, whereas access point did not ($\tau_{\text{AP}} = 0$). Fig. 20 compares the buffering time length for user terminal with the band occupancy when we changed the band occupancy from 0 to 10. The theoretical results suggested that the upper bound was always larger than the lower bound on band occupancy. In addition, Fig. 21 shows the mean waiting times for access point and user terminal. The mean waiting time for access point increased in a linear manner with the buffering time length ($\mathrm{E}[W_{\mathrm{AP}}] = O(\tau_{\mathrm{UT}})$, $\mathrm{E}[W_{\mathrm{UT}}] = O(\tau_{\mathrm{UT}})$). The network system was stable at all buffering time lengths when the buffering time length was zero for access point. This suggests that we should not set a long buffering time for nodes with a higher traffic load.

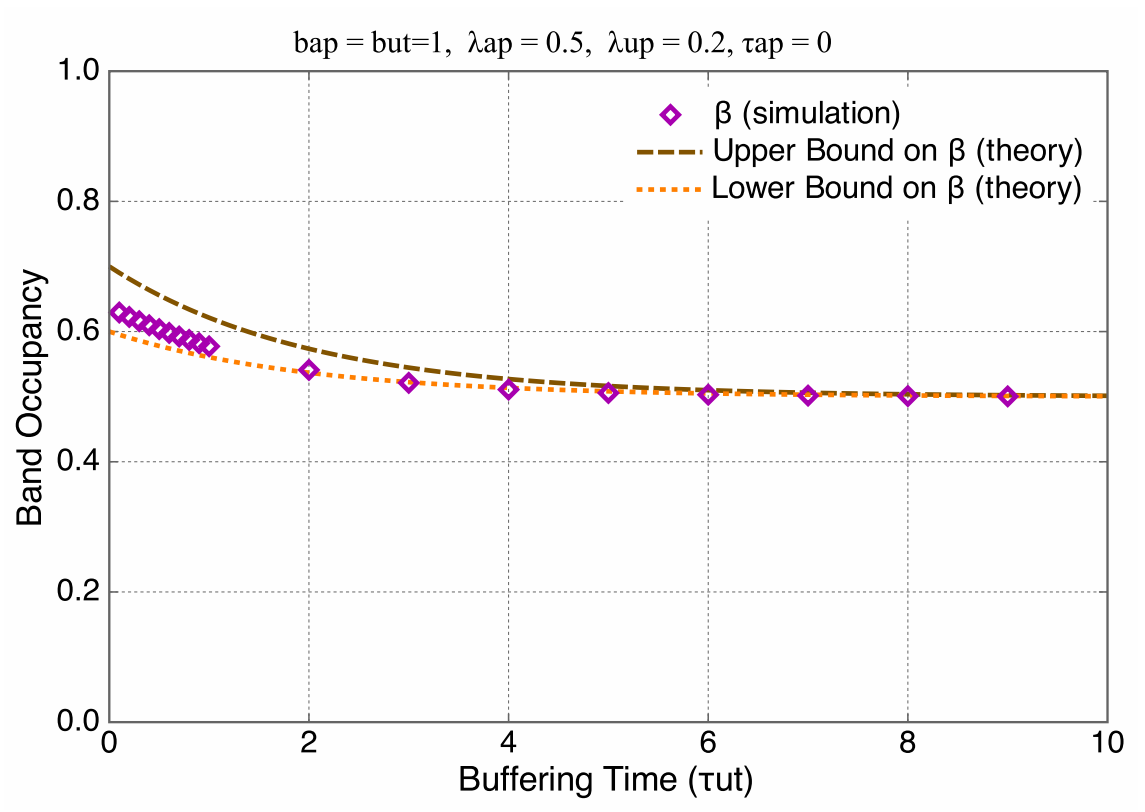

Fig. 20. Buffering Time Length for user terminal vs. Band Occupancy ($\tau_{\mathrm{AP}} = 0$)

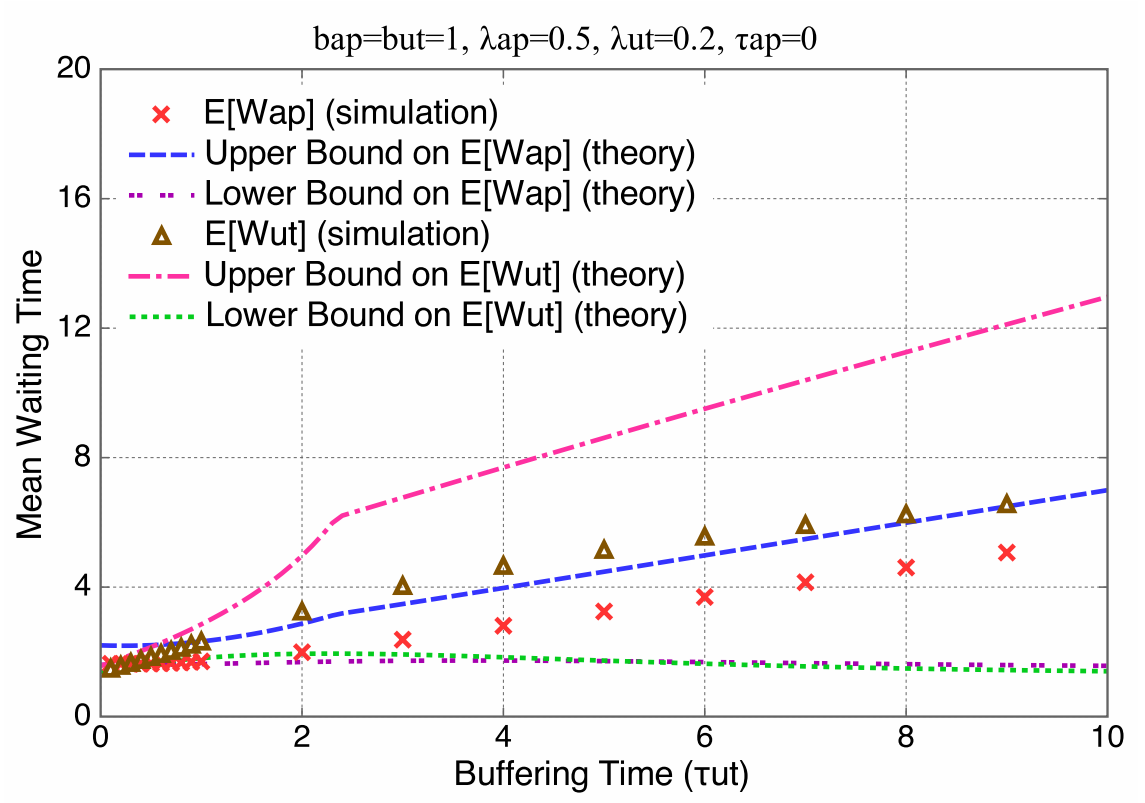

Fig. 21. Buffering Time Length for user terminal vs. Mean Waiting Time ($\tau_{\mathrm{AP}} = 0$)

## VI. Conclusion

In this study, we provided the first theoretical analysis of band occupancy and the mean waiting time for full-duplex communication under traffic that is not fully buffered based on queueing theory, with the closed-form results. The upper bound and lower bound on band occupancy as well as the mean waiting time were determined theoretically. We also presented the results of simulations of the band occupancy and mean waiting time. The theoretical analysis and simulation results provide guidelines for determining the buffering time length. The basic analysis presented in this study demonstrates how the mean waiting time and band occupancy are affected by the buffering time. The buffering time can reduce the throughput with a smaller memory size for the packets on nodes. However, under asymmetrical traffic, we should not set a long buffering time when the traffic intensity load is higher on certain nodes compared with other nodes. Our results support the development of a full-duplex MAC protocol and device design.

## Appendix A
## Proof of Lemma 6

The mean queue waiting time for access point under half-duplex communication is $\mathrm{E}[W_{\mathrm{AP,HD},q}] = \beta_{\mathrm{HD}}\frac{b_{\mathrm{HD}}}{2} + b_{\mathrm{HD}}\mathrm{E}[L_{\mathrm{AP,HD},q}]$ where $\mathrm{E}[L_{\mathrm{AP,HD},q}]$ is the length of the queue for access point due to the PASTA theorem [20]. Using $\mathrm{E}[L_{\mathrm{AP,HD},q}] = \lambda_{\mathrm{AP}}\mathrm{E}[W_{\mathrm{AP,HD},q}]$ [22], the mean waiting time for access point can be reformulated as $\mathrm{E}[W_{\mathrm{AP,HD}}] = \frac{\beta_{\mathrm{HD}} b_{\mathrm{HD}}}{2(1-\beta_{\mathrm{HD}})} + b_{\mathrm{AP}}$. The mean waiting time for user terminal under half-duplex communication can be formulated in a similar manner.

## Appendix B
## Proof of Lemma 7

The mean queue waiting time for access point under ideal full-duplex communication is $\mathrm{E}[W_{\mathrm{AP,IFD},q}] = \rho_{\mathrm{AP}}\frac{b_{\mathrm{AP}}}{2} + b_{\mathrm{AP}}\mathrm{E}[L_{\mathrm{AP,IFD},q}]$ where $\mathrm{E}[L_{\mathrm{AP,IFD},q}]$ is the length of the queue for access point. The mean waiting time for user terminal under ideal full-duplex communication can be obtained in a similar manner.

## Appendix C
## Proof of Section V-A

The mean waiting time in practical full-duplex communication can be reformulated as follows:

$$\begin{aligned}
\mathrm{E}[W_{\mathrm{AP}}] &< \frac{1}{2\left(1-\rho_{\mathrm{AP}}\right)}\left\{\beta_{\mathrm{UT}}(\beta_{\max})b + \rho_{\mathrm{AP}}b\right\} + \frac{\rho_{\mathrm{AP}}}{\lambda_{\mathrm{AP}}} \\
&= \frac{1}{2\left(1-\rho_{\mathrm{AP}}\right)}\left\{\beta_{\mathrm{UT}}(\beta_{\mathrm{HD}})b + \rho_{\mathrm{AP}}b\right\} + b \\
&= \frac{1}{2\left(1-\rho_{\mathrm{AP}}\right)}\beta_{\mathrm{HD}}b + b
\end{aligned}$$

$$< \frac{1}{2\left(1-\beta_{\mathrm{HD}}\right)}\beta_{\mathrm{HD}}b + b = \mathrm{E}[W_{\mathrm{AP,HD}}].$$

Thus, the mean waiting time in practical full-duplex communication is less than that in half-duplex communication.

In addition, the mean waiting time in practical full-duplex communication can be reformulated as follows:

$$\begin{aligned} \mathrm{E}[W_{\mathrm{AP}}] &> \frac{1}{2\left(1-\rho_{\mathrm{AP}}\right)}\left\{\beta_{\mathrm{UT}}(\beta_{\min})b + \rho_{\mathrm{AP}}b\right\} + \frac{\rho_{\mathrm{AP}}}{\lambda_{\mathrm{AP}}} \\ &> \frac{1}{2\left(1-\rho_{\mathrm{AP}}\right)}\rho_{\mathrm{AP}}b + b = \mathrm{E}[W_{\mathrm{AP,IFD}}]. \end{aligned}$$

The mean waiting time in practical full-duplex communication is always greater than that in ideal full-duplex communication.

## Acknowledgment

This study was supported by JSPS KAKENHI Grant Number JP16H01718.